\documentclass{article}
\usepackage[a4paper,top=2.8cm,bottom=2.8cm,left=2.8cm,right=2.8cm]{geometry}
\usepackage{amsmath}
\usepackage{authblk}
  \usepackage{paralist}
  \usepackage{graphics}
  \usepackage{epsfig} 
\usepackage{graphicx}  
\usepackage{epstopdf}
\usepackage{amsmath,amssymb,amsfonts,amsthm}
\usepackage{mathtools}
\mathtoolsset{showonlyrefs}
\usepackage[colorlinks=true]{hyperref}
\hypersetup{urlcolor=blue, linkcolor=red, citecolor=red}
\usepackage{csquotes}
\usepackage{mathtools}
\mathtoolsset{showonlyrefs}
\usepackage{amsfonts}
\usepackage{amssymb}
\usepackage{amsthm}
\usepackage{mathrsfs}
\usepackage{stackrel,graphicx}
\usepackage{enumitem}
\usepackage{cases}
\usepackage{booktabs}
\usepackage{braket}
\usepackage[strict]{changepage}
\usepackage{subfigure}
\usepackage[small,justification=justified]{caption}
\usepackage{titlesec} 
\usepackage{multicol}
\usepackage{pgfplots}
\usepackage[cbgreek]{textgreek}
\usepackage{bbm}
\usepackage{cancel}
\usepackage{multirow}
\usepackage[table]{xcolor}
\usepackage[right]{lineno}

\usepackage[backend=bibtex8,doi=false,eprint=false,giveninits=true,isbn=false,style=ieee,url=false,maxnames=99]{biblatex}
\makeatletter
\def\blx@maxline{77}
\makeatother
\bibliography{bibl_modeling.bib}
\DeclareRedundantLanguages{english,german,french}{english,german,ngerman,french}

\newtheorem*{rem*}{Remark}
\theoremstyle{definition}

\newcommand{\R}{\mathbb{R}}

\providecommand{\keywords}[1]{\textbf{\textit{Keywords---}} #1}

\newenvironment{sistem}
{\left\lbrace\begin{array}{@{}l@{}}}
{\end{array}\right.}

\begin{document}

\title{Randomness-aware multiscale models of glioma invasion and treatment}
\author[1]{Martina Conte\thanks{\texttt{Corresponding author: martina.conte@polito.it}}}
\author[2]{Sandesh Hiremath}
\author[2]{Christina Surulescu}
\affil[1]{\centerline{\small Department of Mathematical Sciences "G.L. Lagrange", Politecnico di Torino} \newline \centerline{\small  Corso Duca degli Abruzzi, 24 - 10129 Torino, Italy}}
\affil[2]{\centerline {\small Department of Mathematics, RPTU Kaiserslautern-Landau} \newline \centerline{\small Gottlieb-Daimler-Str. 48 - 67663 Kaiserslautern, Germany}}

\date{\today}                     
\setcounter{Maxaffil}{0}
\renewcommand\Affilfont{\itshape\small}
\maketitle

\begin{abstract}
In this work, we develop a stochastic multiscale model for glioma growth and invasion in the brain, incorporating the effects of therapeutic interventions. The model accounts for tumor cell migration influenced by brain tissue heterogeneity and anti-crowding mechanisms, while explicitly addressing treatment-related uncertainties through stochastic processes. Starting from a microscopic description of individual cell dynamics, we derive the corresponding system of macroscopic random reaction-diffusion-taxis equations governing cell density and tissue evolution.
Finally, we conduct several numerical experiments to assess the efficacy of different treatment protocols, evaluated with respect to well-established clinical criteria and measurable outcomes.

\end{abstract}

\keywords{Glioma multiscale modeling, Kinetic transport equations, Random reaction-diffusion-taxis equations, Radiotherapy, Tumor control probability}

\section{Introduction}
Gliomas are among the most aggressive and infiltrative brain tumors, characterized by complex growth patterns arising from dynamic interactions between tumor cells and their surrounding microenvironment~\cite{wrensch2002epidemiology,ostrom2014epidemiology}. In particular, gliomas exhibit rapid proliferation and extensive infiltration into healthy brain tissue, which significantly complicates both diagnosis and treatment. Their highly diffusive invasion leads to poorly defined tumor margins, hindering accurate delineation and limiting the extent of safe surgical resection. As a consequence, complete removal of the tumor is rarely possible, and the prognosis for patients remains poor~\cite{wick2018treatment,berens1999those}.

The first-line treatment for newly diagnosed gliomas typically involves surgical resection of the tumor. However, complete resection is often not feasible because the deeply infiltrative tumor rims cannot be fully characterized with standard medical imaging. For this reason, surgery is frequently combined with additional therapies. Among these, radiotherapy is routinely employed to reduce tumor burden (sometimes even before surgery). {The standard radiotherapy protocol for gliomas generally consists of a 6-week course of daily irradiation delivered according to a 5-days-on, 2-days-off schedule, accounting for the weekend, with a total dose of 60 Gy delivered to the identified regions of interest. However,} a variety of alternative radiotherapy protocols have also been investigated. Among the most widely studied are hyper-fractionated~\cite{gondi2021radiotherapy,nieder1999hyperfractionated} and hypo-fractionated~\cite{beckham2024hypofractionated,trone2020survival} regimes, which differ from the standard protocol in the daily radiation dose delivered to the patient and the resulting overall duration of the treatment.

A crucial aspect of all radiotherapy protocols is the presence of substantial uncertainty, which originates from multiple sources. Some of it is due to tumor biology and its microenvironment, including intrinsic variability in radiosensitivity across patients and even within a single tumor, spatial heterogeneity of tissue structure, or fluctuations in nutrient and oxygen availability~\cite{junttila2013influence,telarovic2021interfering}. Other uncertainties originate from treatment delivery and clinical practice, such as patient-positioning variability, brain shift or fluid accumulation during therapy, {variations in the treatment schedule due to missed or delayed treatment sessions,} and limitations in imaging resolution or tumor segmentation, among many others~\cite{segedin2016uncertainties,winey2014geometric}. {The characterization of these sources of uncertainty and the quantification of their individual contributions to treatment response remain challenging, as their effects are often interdependent and difficult to disentangle. A comprehensive characterization of such variability is therefore essential to improve the predictive reliability of radiotherapy models and to better understand its impact on glioma progression and treatment outcomes.}

\paragraph{State of the art}
Mathematical models of tumor growth and invasion have traditionally been formulated through deterministic partial differential equations (PDEs), which describe cell proliferation, diffusion- and taxis-driven migration, along with further interactions with the surrounding tissue. These models have provided insights into glioma invasion, tissue degradation, and the influence of the microenvironment (e.g., see~\cite{conte2021modeling,rockne2009mathematical,swanson2000quantitative,swanson2011quantifying,harpold2007evolution,liang2023bayesian}). However, deterministic frameworks assume that all processes evolve in a predictable manner, which is incompatible with the biological variability observed in clinical settings. Fluctuations in cell motility, tissue properties, and treatment delivery naturally motivate the development of models capable of explicitly incorporating randomness and stochasticity. 

In recent years, partial differential equations with randomness have been introduced to incorporate stochasticity into continuum tumor models. These approaches have been used to describe random perturbations in cell motility, tumor–tissue interactions, and microenvironmental fields (e.g., see~\cite{hiremath2018coupled,hiremath2018mathematical,hiremath2016stochastic,hiremath2015stochastic,Katsaounis2023,Lima2014,Essarrout2022,Mansour2025,Jerez2019}). Such models capture spatial variability in tumor fronts and give rise to more realistic invasive patterns, better reflecting the irregular morphology commonly observed in infiltrative cancers such as gliomas. Notwithstanding these advances, the explicit integration of stochasticity into radiotherapy models has been relatively rarely addressed, in spite of its clinical relevance. Classical mathematical models, including those based on the Linear–Quadratic (LQ) formalism~\cite{fowler1989linear}, typically treat radiobiological processes deterministically. Yet both experimental and clinical evidence demonstrate substantial variability in radiation response. Only a limited number of recent works have begun to introduce stochastic elements into radiotherapy modeling~\cite{kim2012stochastic,saberian2016theoretical,fornalski2011stochastic}. These contributions demonstrate that randomness can significantly alter therapeutic predictions, affecting tumor control probability, relapse timing, or triggering the emergence of resistant subpopulations. Nonetheless, most existing models introduce randomness in a phenomenological manner, lacking a systematic multiscale derivation linking microscopic cellular processes to macroscopic tumor evolution.

To capture the multiscale nature of tumor progression, several modeling frameworks have been proposed based on microscopic and/or mesoscopic descriptions of cell behavior. Many of these are formulated within the framework of the kinetic theory of active particles (KTAPs)~\cite{bellomo2022towards,Bell-KTAP}, which enables a multiscale representation of multicellular systems. Models within this context characterize cell dynamics at the mesoscopic level, beginning from processes occurring at the scale of individual cells. Through well-established asymptotic techniques, based on diffusive or hyperbolic scalings, macroscopic equations are derived from these lower-scale descriptions. In the context of glioma modeling, a variety of systems have been proposed based on this approach~\cite{conte2023mathematical,conte_surulescu2020,conte2020glioma,engwer2015glioma,engwer2016effective,,buckwar2023stochastic,engwer2016multiscale,corbin2018higher,corbin2021modeling,hunt2017multiscale,kumar2020flux,kumar2021multiscale,dietrich2020,painter2013mathematical}. Among these, \cite{conte2020glioma,conte2023mathematical,hunt2017multiscale} investigate tumor evolution in response to different therapeutic interventions,  accounting for their distinct effects on both tumor cells and healthy tissue. However, models that explicitly incorporate multiple therapy approaches remain scarce, and those that directly account for randomness in treatment are even fewer. This gap highlights the need for multiscale PDE frameworks in which stochasticity explicitly enters the radiotherapy terms and propagates through to clinically relevant population-level dynamics. Such models provide a mathematical foundation for randomness-aware predictions of treatment outcomes, facilitate the evaluation of the robustness of treatment protocols, and support the optimization of radiotherapy under realistic variability. The present work contributes towards this aim.
\vspace{0.3cm}

In this work, we develop a multiscale stochastic model for glioma progression and therapy that integrates microscopic, mesoscopic, and macroscopic levels of description using tools from the kinetic theory of active particles. The primary objective is to connect the dynamics of glioma cells and healthy tissue with their responses to therapeutic interventions under inherent randomness. The work is organized as follows.  Section~\ref{sec:modeling} presents the formulation of the model at the subcellular and mesoscopic scales and derives the corresponding macroscopic random PDE system. In Section~\ref{Sec:coef}, we define the coefficient functions and the initial experimental setup used in the numerical simulations. Section~\ref{Sec:NumTest} presents the simulation results, evaluating therapy outcomes under various clinically inspired metrics. We conclude with a discussion of the results in Section~\ref{sec:discussion}. Additional figures are provided in the Supplementary Material.

\section{Model description}\label{sec:modeling}
This section is dedicated to the development of the stochastic model announced above, with the main goal of characterizing the space-time interactions of glioma cells with surrounding normal tissue and assessing their response to different treatment protocols under inherent randomness. In the model, glioma cells migrate preferentially along the anisotropic structure of brain tissue while avoiding highly crowded regions. Cell proliferation depends on the availability of surrounding healthy tissue, which supports tumor growth and provides essential supplies. The effects of therapy account for radiation-induced damage to both tumor and normal tissue, though with different sensitivities, and are modulated through a stochastic process of the Ornstein–Uhlenbeck type. This framework enables the assessment of radiotherapy outcomes using the concepts of Tumor Control Probability (TCP), Normal Tissue Complication Probability (NTCP), and Uncomplicated Tumor Control Probability (UTCP), which respectively quantify the probabilities of effective tumor eradication (TCP), normal tissue damage (NTCP), and of achieving tumor control without inducing severe normal tissue complications (UTCP).

Starting from the microscopic description of interactions between cells and tissue, we formulate the corresponding kinetic transport equations (KTEs) for glioma cells. By performing a formal parabolic scaling limit, we derive the macroscopic random reaction–advection–diffusion equation governing the dynamics of the tumor cell population. Finally, we couple this equation with the ordinary differential equation characterizing the evolution of the healthy tissue, modeled directly at the macroscopic level.

\newcommand{\brho}{\boldsymbol{\varrho}}
\newcommand{\vrhoone}{{^1\!\varrho}}
\newcommand{\vrhotwo}{{^2\!\varrho}}
\subsection{Microscopic scale}\label{micro_section}
On the microscopic scale, we describe the interactions between glioma cells, tissue fibers, and the surrounding cellular environment. We assume that cells exchange information with their microenvironment through various types of cell surface receptors. Specifically, the binding of these transmembrane units is considered to trigger intracellular processes that lead to migration and proliferation. This modeling approach follows the ideas introduced in \cite{kelkel2012multiscale,lorenz2014class} for constructing a multiscale (micro–meso) model of tumor invasion incorporating different types of tactic responses, and subsequently refined in several later works (see \cite{engwer2015glioma,engwer2016effective,engwer2016multiscale,conte_surulescu2020,conte2023mathematical,chiari2025multi,corbin2018higher,corbin2021modeling,hunt2017multiscale} and references therein). In particular, this framework enables the inclusion of both single-cell dynamics, which usually appear in the KTEs, and subcellular processes, which have a significant impact on overall tumor progression.

Given the variables for time $t>0$, spatial position $x\in \mathbb{R}^d$, and microscopic cell velocity $v\in V = s\mathbb{S}^{d-1}$ (namely $v = s\theta$ and $\theta \in \mathbb{S}^{d-1}$, with $\mathbb{S}^{d-1}$ unit sphere in $\mathbb{R}^d$), we denote by $y_1(t)$ the amount of cell receptors bound to tissue fibers and by $y_2(t)$ the amount of transmembrane entities mediating adhesion between tumor cells. The corresponding binding dynamics are governed by simple mass-action kinetics:
\begin{equation}
\begin{split}
&(\bar{R}_0-y_1)+Q\stackrel[k^-_1]{k^+_1}{\rightleftharpoons}y_1\\[0.3cm]
&(\bar{R}_0-y_2)+M\stackrel[k^-_2]{k^+_2}{\rightleftharpoons}y_2\,,
\end{split}
\end{equation}
where $\bar{R}_0$ denotes the total number of receptors on the cell membrane, assumed constant for simplicity. Here, $Q(t,x)$ represents the macroscopic volume fraction of available tissue, irrespective of the orientation of fibers, and $M(t,x)$ the macroscopic tumor cell density. Accordingly, the microscopic receptor-binding dynamics are described by the following system of ordinary differential equations
\begin{equation}\label{micro_eq}
\begin{split}
&\dot{y}_1=k^+_1 Q S_Q(d_r,\brho_Q)(\bar{R}_0-y_1)-k^-_1y_1\\[0.2cm]
&\dot{y}_2=k^+_2 M S_M(d_r,\brho_M)(\bar{R}_0-y_2)-k^-_2y_2\,,
\end{split}
\end{equation}
where $k_1^+$ and $k_1^-$ denote the attachment and detachment rates between tumor cells and tissue fibers, respectively, while $k_2^+$ and $k_2^-$ correspond to the analogous rates for cell–cell interactions. 
The factor 
\begin{equation}\label{Si_ter}
    S_i(d_r, {\brho_i}) := \exp(-\vrhoone_i d_r - \vrhotwo_i d_r^2), \quad {\brho_i = [\vrhoone_i, \vrhotwo_i]^{\top}}
\end{equation}
for $i \in \{ Q, M \}$, models the surviving fraction of cells (tumor or healthy) following a radiation dose $d_r$. This formulation corresponds to the well-known linear–quadratic model. Each population affected by radiotherapy is characterized by specific parameters $\vrhoone$ and $\vrhotwo$, which quantify lethal lesions caused by single radiation tracks and by the interaction of two tracks, respectively. A key parameter in this framework is the ratio $\vrhoone / \vrhotwo$, which reflects the cell’s radiosensitivity and correlates with the cell cycle length: late-responding cells with slower proliferation exhibit smaller $\vrhoone / \vrhotwo$ ratios, whereas early-responding, highly aggressive cells display larger ones. The parameter vectors $\brho_M$ and $\brho_Q$ denote the effects of radiotherapy on tumor and healthy cells, respectively. In what follows, we refer to $S_M$ and $S_Q$ as the radiotherapy survival fractions corresponding to $M$ and $Q$, without explicitly indicating their dependence on $\brho$ and $d_r$.

Considering \eqref{micro_eq}, if we assume that $k_1^-=k_2^-=k^-$, then we define $y(t)=y_1(t)+y_2(t)$ as the total concentration of bounded receptors on the cell membrane, and we lump together the two equations into 
\begin{equation}\label{microLump_eq}
\dot{y}=\left(k^+_1 S_Q Q+k^+_2 S_M M\right)\left(\bar{R}_0 -y\right)-k^-y\,.
\end{equation}
By rescaling $y/{\bar{R}_0}\leadsto y$, the unique steady state of equation \eqref{microLump_eq} reads
\begin{equation}\label{ss_micro}
y^*=\dfrac{B(M,Q)-k^-}{B(M,Q)}
\end{equation}
with 
\begin{equation}\label{def:B_MQ}
B(M,Q):=k^+_1 S_Q Q+k^+_2 S_M M+k^-\,.
\end{equation}

We assume that tumor cells tend to migrate along the gradients of tissue fibers while avoiding overcrowded regions. Consequently, we look at the path of a single cell initially located at position $x_0$ and moving toward position $x$ with velocity $v$ within the density fields $Q$ and $M$, such that $Q(x) = Q(x_0 + v t)$ and $M(x) = M(x_0 - v t)$. Denoting by $z:= y^* - y$ the deviation of $y$ from its steady state $y^*$, {we have $\dot z=\dot{y}^*-\dot y$. From the expression of $y^*$ in~\eqref{ss_micro}, its derivative along the cell trajectory is
\[
\dot y^*=\frac{k^-}{B(M,Q)^2}\dot B(M,Q).
\]
Using the definition of $B(M,Q)$ given in~\eqref{def:B_MQ} and applying the chain rule along the cell trajectory, we obtain
\begin{equation*}
\dot B=k_1^+S_Q\left(v\cdot\nabla_xQ+\partial_tQ\right)+k_2^+S_M\left(-v\cdot\nabla_xM+\partial_tM\right).
\end{equation*}
Combining this expression with~\eqref{microLump_eq}, and using $z=y{^*}-y$, yields}
\begin{equation}
\dot{z}=-zB(M,Q)+\dfrac{k^-}{B(M,Q)^2}\left[k_1^+S_Q \left(v\cdot \nabla_x Q+\partial_t Q\right)-k_2^+S_M \left(v\cdot \nabla_x M-\partial_t M\right)\right]=:G(z,Q,M)\,.
\end{equation}

\begin{rem*}
We remark that the macroscopic densities $Q$ and $M$ depend also on the stochastic realizations $\omega \in \Omega$, hence they are in fact nonnegative stochastic processes defined on $\R_+\times \Omega$. The whole problem is actually set within the framework of a probability space $(\Omega, \mathcal{F}, \mathbb{P})$, with $\mathcal{F}$ being a filtration and $\mathbb{P}$ the underlying probability measure. These stochastic realizations originate from the randomness contained in the radiotherapy term and therefore propagate up to the macroscopic level. For the sake of clarity, we omit here the explicit dependence on $\omega $; a detailed handling will be provided in the forthcoming work \cite{Hiremath2026}.
\end{rem*}

\subsection{Mesoscopic scale}\label{sec2:mesodescription}
We model the mesoscale dynamics of glioma cells using kinetic transport equations that describe velocity-jump processes while incorporating microscopic behavior. Specifically, we consider the mesoscopic cell density function $p=p(t,x,v,y)$ which depends on time $t>0$, spatial position $x\in\mathbb{R}^d$, microscopic velocity $v\in V=s\mathbb{S}^{d-1}$, and state variable $y\in Y=(0,1)$. The choice $v\in V=s\mathbb{S}^{d-1}$ corresponds to tumor cells moving at a constant speed $s>0$. This choice means that we assume constant speeds $ s> 0$ for tumor cells. Following \cite{engwer2015glioma,engwer2016multiscale,conte2020glioma}, we work with the variable $z=y^*-y\in Z\subseteq(y^*-1,y^*)$ rather then $y$ itself. The evolution of the glioma cell density is governed by the kinetic transport equation
\begin{equation}\label{KTE}
\partial_t p +\nabla_x\cdot\,(v p) +\partial_z(G(z,Q,M)p)=\mathcal{L}[\lambda(z)]p+\mathcal{P}\,p-R_M(d_r,\xi_t)p\,.
\end{equation}
Here $\mathcal{L}[\lambda(z)]p$ denotes the turning operator, which models changes in cell velocity due to contact guidance, namely the tendency of glioma cells to align with brain tissue anisotropy, primarily along white matter tracts. This operator has a Boltzmann-type integral form:  
\begin{equation}
\mathcal{L}[\lambda(z)]p:=-\lambda(z)p+\lambda(z)\int\limits_V K(x,v)\,p(t,x,v',z)\,dv'\,.
\end{equation}
The turning rate $\lambda(z):=\lambda_0-\lambda_1z\ge 0$ depends on the microscopic state $z$, where $\lambda_0>0$ and $\lambda_1>0$ represent, respectively, the basal turning frequency and the sensitivity to environmental cues. The integral term describes the reorientation of cells from any previous velocity $v'$ to a new velocity $v$ after interacting with the surrounding tissue. The turning kernel $K(x,v)$ is assumed to be independent of $v'$ and characterizes the dominant directional cue imposed by the orientation of tissue fibers. Following \cite{painter2013mathematical,engwer2015glioma}, we define 
\begin{equation}
K(x,v):=\frac{q(x,\theta)}{\gamma}\,,\qquad \theta=\frac{v}{|v|}\in\mathbb{S}^{d-1}\,,\qquad \gamma=s^{d-1}\,, 
\end{equation}
where $q(x,\theta)$ is the orientational distribution function of tissue fibers, normalized by $\gamma$. This function encodes patient-specific structural information that can be obtained, in clinical practice, via diffusion tensor imaging (DTI). In this study, we assume undirected tissue, i.e., $q(x,\theta)=q(x,-\theta)$ for all $x\in\mathbb{R}^d$. The specific choice of $q(x,\theta)$ will be detailed in Section \ref{Sec:coef}. For convenience, we also define:
\begin{equation}
    \begin{split}
        &\mathbb{E}_q(x):=\int\limits_{\mathbb{S}^{d-1}}\theta q(x,\theta)d\theta\\[0.2cm]
        &\mathbb{V}_q(x):=\int\limits_{\mathbb{S}^{d-1}}(\theta-\mathbb{E}_q(x))\otimes (\theta-\mathbb{E}_q(x))q(x,\theta)d\theta\\[0.2cm]
    \end{split}
\end{equation}
representing the mean fiber orientation and the variance–covariance matrix of the orientation distribution, respectively. The assumed symmetry of $q$ implies $\mathbb{E}_q=0$.

As in \cite{conte2023mathematical,engwer2016effective}, the term $\mathcal{P}\,p$ represents the proliferation process induced by receptor-mediated interactions between glioma cells and brain tissue. We define it as
\begin{equation}
\mathcal{P}\,p:=\mu(M,Q)\int\limits_Z\chi(x,z,z')Q(t,x)p(t,x,v,z')dz'\,.
\end{equation}
Here, proliferation arises as a consequence of cell–tissue interactions. The proliferation rate $\mu(M,Q)$ depends on both the macroscopic density of tumor cells and the macroscopic density of the surrounding brain tissue. In the integral operator, the kernel $\chi(x,z,z')$ characterizes the transition from state $z'$ to state $z$ during proliferation-triggering events occurring at position $x$. No further conditions are required on $\chi$; we only require the nonlinear proliferative operator to be uniformly bounded in the $L^2$-norm, which is biologically reasonable given the spatial constraints that limit cell division.

The last term in \eqref{KTE} models the effects of radiotherapy on the tumor cell population as a function of the delivered radiation dose $d_r$ and the stochastic process $\xi_t$. Specifically, $\xi_t:=\xi(t,\omega)$ models treatment-related uncertainties that affect the efficacy of radiation therapy. In clinical practice, such randomness arises from multiple sources, including variability in dose delivery (e.g., fluctuations in beam intensity, energy, or targeting accuracy), patient-specific factors such as anatomical motion, setup and positioning errors, and biological heterogeneity in cell radiosensitivity. Additional variability can result from environmental and technical noise, including equipment calibration errors or imaging inaccuracies~\cite{segedin2016uncertainties,tanaka2020impact,becksfort2023setup,park2022setup}. {However, disentangling the individual contributions of these factors remains challenging, as their effects may be interdependent and difficult to quantify separately. Consequently, as a first approximation, we assume that} $\xi_t$ acts as a common stochastic driver, affecting the entire tumor cell population and representing the aggregate impact of these random fluctuations on the effective radiation response. Incorporating this stochastic term allows the model to capture variability in treatment outcomes, leading to a more robust and clinically relevant description of tumor evolution under radiotherapy. One natural approach to include such stochastic effects in the model is to define the radiotherapy operator as
\begin{equation}\label{rad_term_gen}
R_M(d_r,\xi_t):=\xi_t\sum_{j=1}^{n} (1-S_M(d_r,\brho_M))f_\delta(t-t_j)
\end{equation}
with $S_M(d_r,\brho_M)$ given in \eqref{Si_ter}. In this formulation, which is reminiscent of its deterministic counterpart~\cite{hunt2017multiscale,conte2023mathematical}, the total dose $d_r$ of the drug is given in smaller fractions to reduce cytotoxicity. Thereby, $n$ is the number of fractions, while $t_j$ denotes the time instants at which ionizing radiation is applied to the patient. For the function $f_\delta(t-t_j)$, we use an exponential distribution centered at $t_j$ with rate parameter $\delta=0.001$. This choice allows us to capture the mild temporal propagation of the radiotherapy effect beyond the effective treatment times $t_j$. To model the stochastic process $\xi_t$, we adopt an Ornstein–Uhlenbeck (OU) process, which is widely used in radiation modeling to describe mean-reverting random fluctuations in treatment efficacy~\cite{oksendal2013stochastic,hlatky1994influence,badri2016optimal}. The OU process satisfies the stochastic differential equation
\begin{equation}\label{Ornstein_proc}
d \xi_t= (\xi_t - \mu_R) dt+ \sigma_R \xi_t dW_t
\end{equation} 
where $W_t$ is a standard Wiener process, $\mu_R$ is the expectation of $\xi_t$, and $\sigma_R$ is the diffusivity parameter controlling the intensity of random fluctuations. 

\subsection{Parabolic scaling of the mesoscopic model}
As clinicians are primarily interested in the macroscopic evolution of the tumor mass, it is convenient to derive effective equations for the macroscopic dynamics of glioma cells, represented by $M$. As noted earlier, the equation for the density of healthy tissue $Q$ is already macroscopic and does not require upscaling. The heuristic derivation presented here follows a classical approach; similar methodologies can be found, e.g., in~\cite{conte2020glioma,chiari2025multi,engwer2016multiscale,dietrich2020,engwer2016effective,hunt2017multiscale,painter2013mathematical,hillen2006m5}. For a rigorous handling of parabolic and hyperbolic upscaling for simplified equations, see~\cite{ZS22}. We perform a rescaling of time and space as $t\to\varepsilon^2 t$ and $x\to\varepsilon x$. In addition, the proliferation term $\mathcal{P}$ and the radiotherapy-induced death term $R_M$ in \eqref{KTE} are scaled by $\varepsilon^2$ to account for mitotic and apoptotic events, which occur on a much longer time scale than cell migration.  {Indeed, the duration of glioma cell cycle is about 57$\pm$ 6 hours \cite{Hoshino1975}, and it typically takes 3-7 days to achieve apoptosis by ionizing radiation~\cite{Murad2018}. On the other hand, the average speed of glioma cells can reach $40-70 \mu m/h$ or even more, depending on the substrate on which they migrate (see, e.g.,~\cite{Monzo2016}).}  The rescaled kinetic transport equation for glioma cells then reads:
\begin{equation}\label{KTE_rescale}
\varepsilon^2\partial_t p +\varepsilon\nabla_x\cdot\,(v p) +\partial_z(G(z,Q,M)p)=\mathcal{L}[\lambda(z)]p+\varepsilon^2\mathcal{P}\,p-\varepsilon^2R_M(d_r,\xi_t)p\,.
\end{equation}
We define the following moments of the distribution function $p$:
\begin{equation*}
\begin{split}
&m(t,x,v)=\int\limits_Zp(t,x,v,z)dz\quad\quad \quad\quad m^z(t,x,v)=\int\limits_Zz\,p(t,x,v,z)dz \\[0.2cm]
& M(t,x)=\int\limits_Vm(t,x,v)dv  \quad\quad \quad \quad\,\,\quad  M^z(t,x)=\int\limits_Vm^z(t,x,v)dv\,.
\end{split}
\end{equation*}
Higher-order moments are neglected based on the assumption that subcellular dynamics occur much faster than events at larger scales, i.e., $z \ll 1$, and that $p$ is compactly supported in the phase space $\mathbb{R}^d\times V\times Z$. Integrating the rescaled equation \eqref{KTE_rescale} with respect to $z$ yields
\begin{equation}
\begin{split}
\varepsilon^2\partial_t m+\varepsilon\nabla_x\cdot(v m)=&-\lambda_0\left(m-\dfrac{q}{\gamma}M\right)+\lambda_1\left(m^z-\dfrac{q}{\gamma}M^z\right)\\[0.1cm]
&+\varepsilon^2\int\limits_Z\mu(M)\int\limits_Z\chi(x,z,z')p(z') Q dz'dz -\varepsilon^2R_M(d_r,\xi_t)m\,.
\end{split}
\end{equation}
Assuming that $\chi(x,z,z')$ is a probability kernel with respect to $z$ for all $(x,z')$, the first moment equation becomes:
\begin{equation}
\partial_t m+\nabla_x\cdot(v m)=-\lambda_0\left(m-\dfrac{q}{\gamma}M\right)+\lambda_1\left(m^z-\dfrac{q}{\gamma}M^z\right)+\mu(M) Q \,m-R_M(d_r,\xi_t)m\,.
\label{meq}
\end{equation}
Multiplying \eqref{KTE_rescale} by $z$ and integrating w.r.t $z$ yields the first-order moment equation in $z$
\begin{equation*}
\begin{split}
\varepsilon^2\partial_t m^z&+\varepsilon\nabla_x\cdot(v m^z)\!+\!\!\int\limits_Zz\partial_z\left[G(z,Q,M)p(z)\right]dz=\!\!\int\limits_Zz\mathcal{L}[\lambda(z)]p(z)dz+\varepsilon^2\mu(Q) Q \,\,m^z\!-\!R_M(d_r,\xi_t)m^z.
\end{split}
\end{equation*}
The calculation of the integral term on the left- and right-hand sides leads to the following equation for $m^z$:
\begin{equation}
\begin{split}
\varepsilon^2\partial_t m^z+\varepsilon\nabla_x\cdot(v m^z)=&-\lambda_0\left(m^z-\dfrac{q}{\gamma}M^z\right)-B(M,Q)m^z+\varepsilon^2\mu(M)Q\,m^z-\varepsilon^2R_M(d_r,\xi_t)m^z+\\[0.2cm]
&\dfrac{\varepsilon k^-}{B(M,Q)^2}\left[\left(k_1^+ S_Q \left(v\cdot \nabla_x Q+\varepsilon\partial_t Q \right)-k_2^+ S_M  \left(v\cdot \nabla_x M+\varepsilon\partial_t M\right)\right)\,m\right]\,.
\end{split}
\label{mzeq}
\end{equation}
We then introduce Hilbert expansions for the previously defined moments:
\begin{equation*}
\begin{split}
&m(t,x,v)=\sum_{k=0}^{\infty}\epsilon^km_k\quad\quad \quad\,\, m^z(t,x,v)=\sum_{k=0}^{\infty}\epsilon^km^z_k\\[0.2cm]
&M(t,x)=\sum_{k=0}^{\infty}\epsilon^kM_k  \quad\quad \quad \quad\,\, M^z(t,x)=\sum_{k=0}^{\infty}\epsilon^kM^z_k\,.
\end{split}
\end{equation*}
Similarly, the proliferation rate is expanded as $\mu(M)=\mu(M_0)+O(\epsilon)$, where $M_0$ denotes the leading-order macroscopic tumor density, while
\begin{equation}
    B(M,Q)=B(M_0,Q)+k_2^+ S_M (M-M_0)+O(|M-M_0|^2)
\end{equation}
and
\begin{equation}
    \dfrac{1}{B(M,Q)^2}=\dfrac{1}{B(M_0,Q)^2}-\dfrac{2k_2^+S_M}{B(M_0,Q)^3}(M-M_0)+O(|M-M_0|^2)\,.
\end{equation}
By substituting these expansions into the moment equations \eqref{meq} and \eqref{mzeq}, we can systematically collect terms corresponding to each power of $\varepsilon$. The resulting hierarchy allows us to derive a macroscopic equation for $M_0(t,x)$ that captures the leading-order dynamics of the glioma cell population. Specifically:\\[0.2cm]
\noindent $\epsilon^0$ terms:
\begin{align}
&0=-\lambda_0\left(m_0-\dfrac{q}{\gamma}M_0\right)+\lambda_1\left(m_0^z-\dfrac{q}{\gamma}M_0^z\right)\label{m0eq}\,,\\[0.2cm]
&0=-B(M_0,Q)m_0^z-\lambda_0\left(m_0^z-\dfrac{q}{\gamma}M_0^z\right)\,, \label{m0zeq}
\end{align}

\noindent $\epsilon^1$ terms:
\begin{align}
&\nabla_x\cdot(v m_0)=-\lambda_0\left(m_1-\dfrac{q}{\gamma}M_1\right)+\lambda_1\left(m_1^z-\dfrac{q}{\gamma}M_1^z\right)\,, \label{m1eq}\\[0.3cm]
&\nabla_x\!\cdot\!(v m_0^z)+B(M_0,Q)m_1^z-\!\dfrac{k^-}{B(M_0,Q)^2}\left(k_1^+ S_Q v\cdot \nabla_x Q-k_2^+ S_M v\cdot \nabla_x M_0\right)\, m_0\!=\!-\lambda_0\left(m_1^z-\dfrac{q}{\gamma}M_1^z\right)\,,\label{m1zeq}
\end{align}

\noindent $\epsilon^2$ terms:
\begin{align}
&\partial_tm_0+\nabla_x\cdot(v m_1)=-\lambda_0\left(m_2-\dfrac{q}{\gamma}M_2\right)+\lambda_1\left(m_2^z-\dfrac{q}{\gamma}M_2^z\right)+\mu(M_0) Q m_0-R_M(d_r,\xi_t)m_0.\label{m2eq}
\end{align}
Integrating \eqref{m0zeq} with respect to $v$ we have
\begin{equation}\label{m0zeq_1}
0=-M_0^zB(M_0,Q)\Longrightarrow M_0^z=0\,.
\end{equation}
Considering again equation \eqref{m0zeq} with \eqref{m0zeq_1}, we have
\begin{equation}
0=-(B(M_0,Q)+\lambda_0)m_0^z\quad \Longrightarrow m_0^z=0\,.
\label{m0zeq_2}
\end{equation}
From equation \eqref{m0eq}, using \eqref{m0zeq_1} and \eqref{m0zeq_2}, we obtain:
\begin{equation}
0=-\lambda_0m_0+\lambda_0\dfrac{q}{\gamma}M_0\quad\Longrightarrow m_0=\dfrac{q}{\gamma}M_0\,.
\label{m0eq_1}
\end{equation}

\noindent Considering \eqref{m1zeq} and integrating it with respect to $v$, we have
\begin{equation}\label{m1zeq_1}
\int\limits_V\nabla\cdot (v  m_0^z)dv=-B(M_0,Q)M_1^z+\dfrac{k^-}{B(M_0,Q)^2}\int\limits_V\dfrac{q}{\gamma}M_0\left(k_1^+ S_Q v\cdot \nabla_x Q-k_2^+ S_M v\cdot \nabla_x M_0\right)dv\,.
\end{equation}
In particular, using \eqref{m0zeq_2} and the symmetry assumption on the fiber distribution $q$, equation \eqref{m1zeq_1} becomes:
\begin{equation}
0=-B(M_0,Q)M_1^z\quad \Longrightarrow M_1^z=0\,.
\label{m1zeq_2}
\end{equation}
Considering again equation \eqref{m1zeq} and using \eqref{m1zeq_2} and \eqref{m0zeq_2}, we get
\begin{equation}\label{m1zeq_3}
m_1^z =\dfrac{k^-}{(B(M_0,Q)+\lambda_0)\,B(M_0,Q)^2}\left(k_1^+ S_Q v\cdot \nabla_x Q-k_2^+ S_M v\cdot \nabla_x M_0\right)m_0\,.
\end{equation}
Considering \eqref{m1eq} and using \eqref{m1zeq_2}, we obtain
\begin{equation}\label{m1eq_1}
\nabla_x\cdot(v m_0)=-\lambda_0\left(m_1-\dfrac{q}{\gamma}M_1\right)+\lambda_1m_1^z \Longrightarrow  \bar{\mathcal{L}}[\lambda_0]m_1:=-\lambda_0 m_1+\lambda_0\dfrac{q}{\gamma}M_1=\nabla_x\cdot(v m_0)-\lambda_1m_1^z\,.
\end{equation}
In order to get an expression for $m_1$, we would like to invert the operator $\bar{\mathcal{L}}[\cdot]$. Similar to \cite{painter2013mathematical,engwer2015glioma}, we define it on the weighted $L^2$-space $L^2_q(V)$, in which the measure $dv$ is weighted by $q(x,\theta)/\gamma$. In particular, $L^2_q(V)$ can be decomposed as $L^2_q(V)=<q/\gamma>\oplus<q/\gamma>^\perp$. For the properties of the chosen turning kernel, $\bar{\mathcal{L}}[\cdot]$ is a compact Hilbert–Schmidt operator with kernel $<q/\gamma>$, so that we can compute its pseudo-inverse on $<q/\gamma>^\perp$. Thus, to determine $m_1$ from \eqref{m1eq_1}, we verify the solvability condition:
\begin{equation}
\int\limits_V \left[\nabla_x\cdot(v m_0)-\lambda_1m_1^z \right]dv=0\,,
\end{equation}
which holds due to \eqref{m1zeq_2} and the assumption of undirected tissue. Consequently, we obtain:
\begin{equation}\label{m1eq_3}
m_1=-\dfrac{1}{\lambda_0}\left[\nabla_x\cdot(v m_0)-\dfrac{\lambda_1k^-}{(B(M_0,Q)+\lambda_0)\,B(M_0,Q)^2}\left(k_1^+ S_Q v\cdot \nabla_x Q-k_2^+ S_M v\cdot \nabla_x M_0\right)\,m_0\right]
\end{equation}
and
\begin{equation}
M_1=0\,.
\label{m1eq_2}
\end{equation}
Finally, integrating \eqref{m2eq} with respect to $v$ leads to the macroscopic evolution equation:
\begin{equation}\label{m2eq_1}
\partial_tM_0+\int\limits_V\nabla_x\cdot(v m_1)dv=\mu(M_0)QM_0-R_M(d_r,\xi_t)M_0
\end{equation}
with
\begin{equation}\label{m2eq_2}
\begin{split}
&\int\limits_V\nabla_x\cdot(v m_1)dv=\\[0.3cm]
&\int\limits_V\nabla_x\!\cdot\!\left[v\left(\!-\dfrac{1}{\lambda_0}\left(\nabla_x\!\cdot\!(v m_0)\!-\!\dfrac{\lambda_1k^-}{(B(M_0,Q)+\lambda_0)\,B(M_0,Q)^2}\left(k_1^+ S_Q v\cdot \nabla_x Q-k_2^+ S_M v\cdot \nabla_x M_0\right)\,m_0\right)\right)\right]dv=\\[0.3cm]
&\nabla_x\!\cdot\!\left[\int\limits_V -\dfrac{1}{\lambda_0}v\otimes v\,\nabla_x\!\cdot\!\left(\dfrac{q}{\gamma}M_0\right)\right]dv+\nabla_x\cdot\left[\dfrac{\lambda_1k^-k_1^+S_Q}{\lambda_0(B(M_0,Q)+\lambda_0)B(M_0,Q)^2} \int\limits_V v\otimes v\,\dfrac{q}{\gamma}dv\,M_0\,\nabla_x Q \right]-\\[0.2cm]
&\nabla_x\cdot\left[\dfrac{\lambda_1k^-k_2^+S_M}{\lambda_0(B(M_0,Q)+\lambda_0)B(M_0,Q)^2} \int\limits_V v\otimes v\,\dfrac{q}{\gamma}dv\,M_0\, \nabla_x M_0\right]\,.
\end{split}
\end{equation}
Introducing the effective diffusion tensor
\begin{equation}\label{DT}
\mathbb{D}_M(x):=\dfrac{1}{\lambda_0\gamma}\int\limits_Vv\otimes v\,q(x,\theta)dv=\dfrac{s^2}{\lambda_0}\int\limits_{\mathbb{S}^{d-1}}\theta\otimes\theta\,q(x,\theta)d\theta=s^2\mathbb{V}_q(x)
\end{equation}
and, in view of \eqref{m1eq_2}, neglecting the $\varepsilon$-correction term for $M$ and the higher-order moments, we can write the macroscopic PDE for tumor density $M(t,x)$ as
\begin{equation}\label{M0_eq}
\begin{split}
  &\partial_tM-\nabla_x \nabla_x:\,\left(\mathbb{D}_M(x) M\right)+\nabla_x\cdot \left[\dfrac{\lambda_1k^-}{(B(M,Q)+\lambda_0)B(M,Q)^2}\mathbb{D}_M(x)\,\left(k_1^+ S_Q \nabla_x Q-k_2^+ S_M \nabla_x M\right)\,M\right]\\[0.2cm]
  &=\left(\mu(M) Q -R_M(d_r,\xi_t)\right)M  
\end{split}
\end{equation}
where we recall that 
\begin{equation}
    B(M,Q)=Q S_Q (d_r,\brho_Q)k^+_1+M S_M (d_r,\brho_M)k^+_2+k^-\,.
\end{equation} 
In particular, among the various possible expressions for the growth rate $\mu(M)$, following \cite{engwer2016effective,conte2020glioma}, we adopt a logistic-type formulation to account for the self-limiting nature of the growth process:
\begin{equation}\label{growth_rate}
\mu(M) = \mu_{M}(1 - M)
\end{equation}
where $\mu_{M} > 0$ denotes the constant proliferation rate.

The equation for $M$ is  coupled with the following evolutionary equation for healthy tissue, which is directly stated at the macroscopic level:
\begin{equation}
\partial_t Q=-d_Q \dfrac{M Q}{1+M}-R_Q(d_r,\xi_t)Q\,.
\end{equation}
Here, $d_Q$ is the tumor-mediated tissue degradation rate, and $R_Q(d_r,\xi_t)$ represents the radiotherapy effect on healthy tissue.

Thus, the coupled macroscopic system of stochastic PDEs describing tumor and healthy tissue evolution is:
\begin{equation}
\begin{sistem}
\begin{aligned}
  \partial_tM-&\nabla_x \nabla_x:\,(\mathbb{D}_M(x) M)+\nabla_x\cdot \left[\dfrac{\lambda_1k^-}{(B(M,Q)+\lambda_0)B(M,Q)^2}\mathbb{D}_M(x)\,\left(k_1^+ S_Q \nabla_x Q-k_2^+ S_M \nabla_x M\right)\,M\right]=\\[0.2cm]
  &\left[\mu_{M}(1 - M)Q-R_M(d_r,\xi_t)\right]M  \,,
\end{aligned} \\[1.4cm]
\partial_t Q=-d_Q \dfrac{M Q}{1+M}-R_Q(d_r,\xi_t)Q\,.
\end{sistem}
\label{mac_sis}
\end{equation}
This system has been deduced considering $x\in\mathbb{R}^d$ and it has to be supplemented with adequate initial conditions. For the numerical experiments performed in Section \eqref{Sec:NumTest} we will consider the system to
be set in a bounded, sufficiently regular domain $D\subseteq\mathbb{R}^2$ and endow it with no-flux boundary conditions, which are obtained, e.g., in a similar way to that presented in~\cite{plaza2019derivation}.

\newcommand{\cgam}{\text{\rm \textgamma}}
\newcommand{\clam}{~~\text{\rm \textlambda}}

\section{Assessment of coefficients and set-up of numerical experiments}\label{Sec:coef}
In this section, we specify the coefficients and functions defining system \eqref{mac_sis}, and we describe the {\it in silico} experimental setting used in Section \ref{Sec:NumTest}.

The computation of the tumor diffusion tensor $\mathbb{D}_M(x)$, introduced in \eqref{DT}, requires prescribing a form for the (mesoscopic) orientational distribution function of tissue fibers, $q(x,\theta)$. Various formulations of this function have been proposed in the literature (see, e.g., \cite{conte2020glioma,shyntar2025first}). Following \cite{conte2023mathematical,conte_surulescu2020}, we employ the orientation distribution function (ODF), defined as
\begin{equation}\label{qODF}
q(x,\theta)=\dfrac{1}{4\pi|\mathbb D_W(x)|^{1/2}\left(\theta^\top(\mathbb D_W(x))^{-1}\theta\right)^{3/2}},
\end{equation}
where $\mathbb{D}_W(x)$ denotes the water diffusion tensor obtained from the patient-specific DTI data. For the radiotherapy terms $R_i(d_r,\xi_t)$, with $i = M, Q$, we recall the general definition \eqref{rad_term_gen}, namely
\begin{equation}
    \begin{split}
&R_M(\xi_t) := \xi_t \sum_{j=1}^n \Big(1-S_M(d_r,\brho_M) \Big) f_\delta(t-t_j)\,, \\
&R_Q(\xi_t) := \xi_t \sum_{j=1}^n \Big(1-S_Q(d_r,\brho_Q) \Big) f_\delta(t-t_j)\,.
    \end{split}
\end{equation}
As described in Section \ref{sec2:mesodescription}, for the function $f_\delta(t-t_j)$, we use an exponential distribution with maximum at $t_j$ and having rate parameter $\delta=0.001$. Regarding the radiation dose, we consider a total dose of $d_r = 60\,\mathrm{Gy}$, divided into a varying number of fractions $n$ according to the specific therapeutic protocol. {For simplicity, we assume that the prescribed dose is uniformly distributed, and the possible spatially heterogeneous distribution of the dose within the target and surrounding healthy tissue is not explicitly modeled.} The radiotherapy parameters $\vrhoone_i$ and $\vrhotwo_i$, which differ between tumor cells ($i=M$) and the healthy tissue ($i=Q$) are listed in Table~\ref{parameter}. The growth rate $\mu(M_0)$ has already been introduced in \eqref{growth_rate}. The remaining constant parameters appearing in system \eqref{mac_sis}, together with their ranges and literature sources, are reported in Table \ref{parameter}.

\begin{table} [!h]
\begin{center}
   \begin{tabular}{c|c|c|c} \hline
   \toprule  
   \rule{0pt}{1.5ex}Parameter & Description & Value (unit) & Source \\
  \midrule
   \rule{0pt}{3ex}$\lambda_0$ & basal turning rate &$0.001$ (s$^{-1})$ & \cite{conte_surulescu2020,Sidani}\\[0.5ex]
\rule{0pt}{3ex}$\lambda_1$ & sensitivity to environmental cues &$0.001$ (s$^{-1})$ &\cite{conte_surulescu2020,Sidani}\\[0.5ex]
\rule{0pt}{3ex}$s$ & tumor cells speed  & $[0.0042, 0.0084]\cdot10^{-3}$ (mm $\cdot$ s$^{-1})$ & \cite{conte_surulescu2020,diao2019}\\[0.5ex]
\rule{0pt}{3ex}$k^+_1$ & tumor-ECM attachment rate & $0.034$ (s$^{-1})$&\cite{Lauffenburger,conte_surulescu2020}\\[0.5ex]
\rule{0pt}{3ex}$k^+_2$ & cell-cell attachment rate & $0.034$ (s$^{-1})$&\cite{Lauffenburger}\\[0.5ex]
\rule{0pt}{3ex}$k^-$ & detachment rate & $0.01$ (s$^{-1})$ & \cite{Lauffenburger}\\[0.5ex]
\rule{0pt}{3ex}$\mu_{M}$ & tumor proliferation rate & $[0.11, 9]\cdot10^{-6}$ (s$^{-1})$ & \cite{ke2000,bhatia2024tumor}  \\[0.5ex]
\rule{0pt}{3ex}$\vrhoone_M$ & single radiation track lesion on tumor& $[0.04, 0.11]$ (Gy$^{-1})$ &\cite{qi2006,barazzuol2010}\\[0.5ex]
\rule{0pt}{3ex}$\vrhoone_Q$ & single radiation track lesion on tissue& $0.00025$ (Gy$^{-1})$& \cite{kroos2019}\\[0.5ex]
\rule{0pt}{3ex}$\vrhotwo_M$ & two radiation tracks lesions on tumor  & $[0.006, 0.019]$ (Gy$^{-2})$ &\cite{qi2006,barazzuol2010}\\[0.5ex]
\rule{0pt}{3ex}$\vrhotwo_Q$ &  two radiation tracks lesions on tissue &$0.00005$ (Gy$^{-2})$ &\cite{kroos2019}\\[0.5ex]
\rule{0pt}{3ex}$d_Q$ & tissue degradation rate (by acidity) &$[0.0 6, 8.3]\cdot10^{-7}$ (s$^{-1})$ &\cite{conte_surulescu2020}\\[0.5ex]
\rule{0pt}{3ex}{$\mu_R$} & {expectation of the OU process} & {$1$} &{This work}\\[0.5ex]
\rule{0pt}{3ex}{$\sigma_R$} & {diffusivity of the OU process} &{$0.65$} &{This work}\\[0.5ex]
\bottomrule
    \end{tabular}
\end{center}
\caption{{\bf Model parameters} (dimensional).}
 \label{parameter}
\end{table}

\subsection{Initial setting}
The numerical simulations of the macroscopic system~\eqref{mac_sis} are conducted in a two-dimensional setting to investigate, \textit{in silico}, different scenarios of tumor response to therapy under varying treatment protocols. Simulations are performed using a self-developed MATLAB code (MathWorks, Inc., Natick, MA) and employing the parameters listed in Table \ref{parameter} and the above-defined functions. 
{A Galerkin finite element method based on continuous piecewise-linear Lagrange elements ($P_1$) is used for spatial discretization \cite{thomee2007galerkin}. The computational domain is discretized using a triangular mesh consisting of 41258 nodes and 80466 elements. Time integration is performed using an IMEX Euler scheme \cite{ascher2008numerical} with the time step $\Delta t=1$ day,} treating the diffusion term implicitly and the drift and reaction terms explicitly. For each treatment protocol, 200 stochastic realizations are computed to account for variability in the simulations.

{For the computational domain $D \subseteq \mathbb{R}^2$, we consider} a horizontal brain slice reconstructed from MRI data. {In what follows, $(x,y)\in D\subseteq\mathbb{R}^2$ denotes a point in the domain $D$.} The macroscopic tensor $\mathbb{D}_M(x,y)$ is precomputed using DTI data and the ODF defined in \eqref{qODF}. The {MRI and DTI datasets were} acquired at the Hospital Galdakao-Usansolo (Galdakao, Spain) and approved by its Ethics Committee; all procedures adhered to institutional guidelines. 

\begin{figure}[h!]
       \centering
       \includegraphics[width=\linewidth]{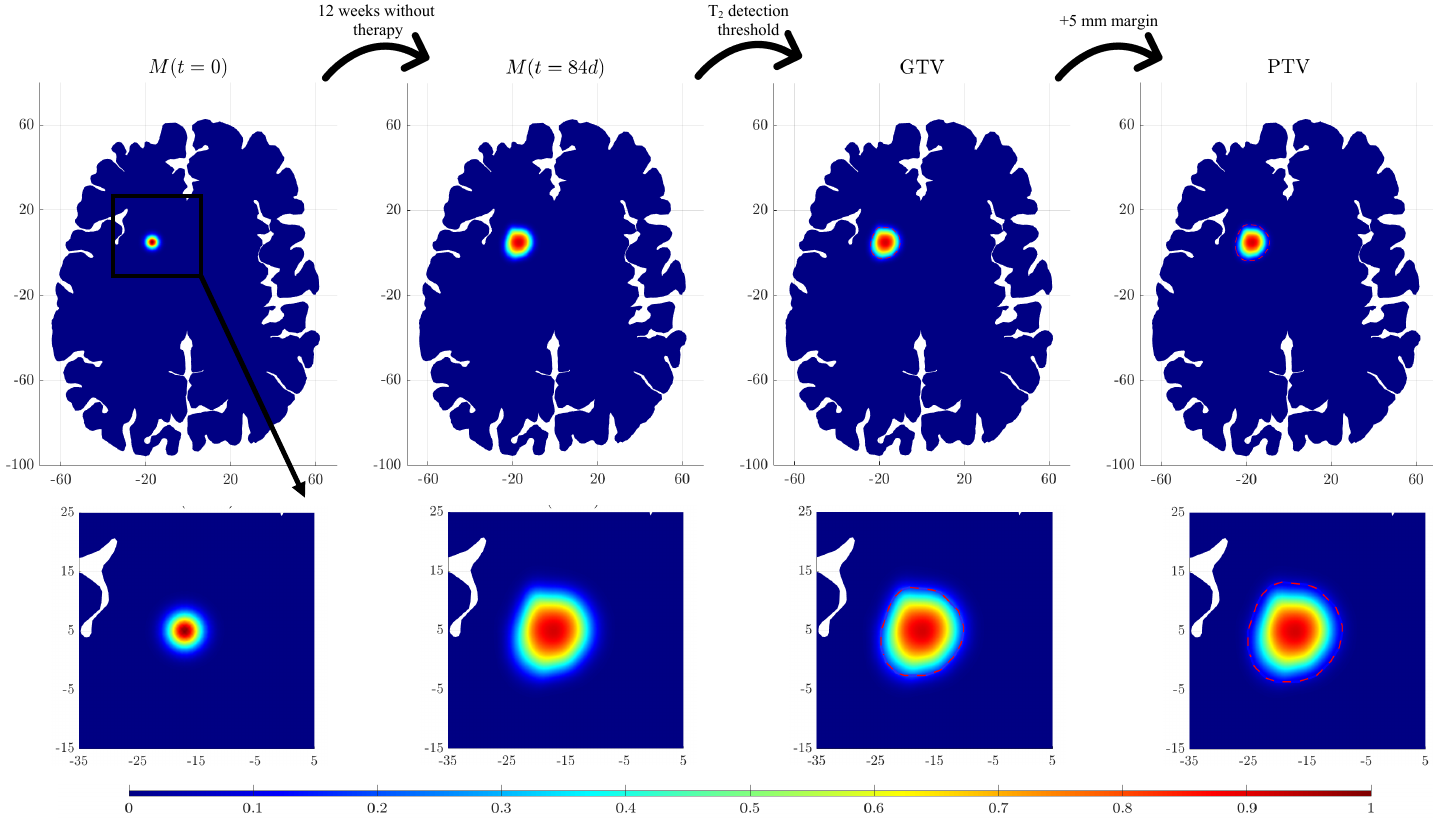}
       \caption{{\bf Definition of the initial setting.} Schematic representation of the model initialization, including the radiotherapy planning volumes. The second row shows magnifications of the main area interested by the tumor mass.}
       \label{Scheeme_InCon}
\end{figure}

The initial conditions are established in two stages. First, we simulate tumor growth in the absence of therapy to generate baseline conditions. Specifically, we initialize a Gaussian-like aggregate of tumor cells centered at $(x_{0,M}, y_{0,M}) = (-17, 5)$ within the right hemisphere of the brain slice representing our computational domain $D \subseteq \mathbb{R}^2$:
\begin{equation}\label{IC_M}
    M_0(x,y)=e^{\frac{(x-x_{0,M})^2+(y-y_{0,M})^2)}{8}}.
\end{equation}
For the initial distribution of normal tissue, we adopt the formulation from \cite{engwer2016effective}, i.e.,
\begin{equation}\label{IC_Q}
    Q_0(x,y)=1-\dfrac{l_c^3(x,y)}{h^3},
\end{equation}
where $h$ denotes the voxel side length of the DTI dataset and $l_c$ is a characteristic length estimated as
\begin{equation}
    l_c(x,y)=\sqrt{\dfrac{~~\text{\rm tr}(\mathbb{D}_W(x,y))h^2}{4l_1}},
\end{equation}
with $l_1$ being the leading eigenvalue of the diffusion tensor $\mathbb{D}_W(x,y)$. Here, we use $h = 0.875$ mm, while $l_1$ is directly estimated from the DTI data in each voxel. To define the initial conditions for treatment simulations, the simulated tumor is first let to evolve without therapeutic intervention for 12 weeks. The resulting tumor distribution at the end of this period serves as the initial condition for treatment assessment. Specifically, in radiotherapy planning, three main volumes are defined \cite{burnet2004defining}. The {\it Gross Tumor Volume} (GTV) represents the tumor visible on radiological images, where the cancer cell density is highest. The {\it Clinical Target Volume} (CTV), which cannot be fully imaged, encompasses the GTV plus a margin (typically 2–3 mm) to include potential microscopic disease spread. For example, for high-grade gliomas it is standard practice to apply a uniform margin around the gross tumor to account for microscopic infiltration, regardless of individual considerations. Finally, the {\it Planning Target Volume} (PTV) extends the CTV by an additional margin (typically 5–10 mm in total from the GTV) to account for geometric and treatment delivery uncertainties. In particular, this is a geometric concept designed to ensure that the radiotherapy dose is actually delivered to the CTV. To replicate clinical practice as closely as possible, we define the GTV at the end of the 12-week untreated growth phase. The GTV corresponds to the imaged tumor region, {defined by assuming an imaging detection threshold of 16\% of the tumor carrying capacity, based on the detection limit associated with T2-Gd MRI as reported in \cite{swanson2011quantifying}. This threshold might be itself subject to uncertainty, as the correspondence between MRI-defined tumor regions and the underlying tumor cell density may vary across patients and brain locations~\cite{lipkova2019personalized}.} The PTV is then obtained by expanding the GTV uniformly by 5 mm in all directions. This PTV definition is subsequently used to evaluate treatment efficacy metrics. A schematic illustration of this procedure is provided in Figure \ref{Scheeme_InCon}.

\section{Numerical experiments and evaluation metrics}\label{Sec:NumTest}
{In this section, we present the results obtained from the numerical simulations, analyzing the outcomes of a range of treatment schedules derived from different fractionation protocols using multiple evaluation metrics. Specifically, for each treatment protocol, 200 independent stochastic realizations of glioma cell growth and response to therapy are performed over the prescribed time interval. The same set of simulations is then analyzed using the different metrics described in the following subsections, each focusing on specific aspects of treatment response and evaluated in the relevant regions and at the appropriate time points. As mentioned in the introduction,} the standard radiation protocol is defined as a 6-week course following a 5-days-on, 2-days-off schedule (to account for weekends), delivering a total dose of 60 Gy. For simplicity, all simulations begin on a Monday, allowing five consecutive treatment days before the two-day rest period. The number of treatment days, denoted as $D_T$, represents the days on which radiation is actually administered; for the standard protocol, $D_T = 30$.

In clinical settings, radiation is delivered in a full three-dimensional volume; however, in this study we restrict our analysis to two-dimensional cross-sections. Radiotherapy aims to maximize tumor cell kill while minimizing normal tissue toxicity, typically employing daily doses between 1 and 2.5 Gy. Hyper- and hypo-fractionation modify the standard protocol by adjusting dose per fraction and treatment frequency, respectively \cite{o1997response}. Although clinical thresholds for radiation toxicity and necrosis are well established \cite{abou2004theoretical}, we initially neglect these limits to isolate the effects of accelerated and decelerated schedules on a virtual tumor. This simplification is justified, as techniques such as Gamma Knife radiosurgery \cite{gerosa2003role} can deliver more than 50 Gy in a single session, deliberately trading increased toxicity risk for enhanced local control.

Several fractionation schemes were simulated, corresponding to total treatment durations of ${D_T = 15, 20, 25, 30, 35, 40}$, and 45 days \cite{o1997response, wheldon1998linear, rockne2009mathematical}. In all cases, the maximum total dose was held constant at 60 Gy.

\subsection{Impact of detection thresholds on tumor volume estimation}\label{subsec:TV}
As a first numerical experiment, we analyze the effects of the different treatment protocols introduced above, as well as the impact of the detection threshold on their assessment. Clinically, glioma growth and progression are primarily monitored through MRI imaging \cite{nelson2003imaging}. Specifically, analyses commonly rely on gadolinium-enhanced T1-weighted (T1Gd) and T2-weighted MRI modalities, which differ in their depiction of tissue contrast. T1Gd images more clearly delineate the bulk of the enhancing lesion, whereas T2-weighted images capture both edema and more diffuse tumor infiltration, albeit at a much lower detection threshold. However, all clinical imaging techniques are limited by their sensitivity thresholds, whereas it is well established that glioma cells infiltrate far beyond regions of visible abnormality. We therefore exploit the advantage of {\it in silico} simulations of radiotherapy, {which allow us to track the full simulated tumor distribution beyond the clinical detection threshold and to compare this idealized tumor burden with the portion that would be detectable through clinical imaging.}

\begin{figure}[h!]
       \centering
       \includegraphics[width=.8\linewidth]{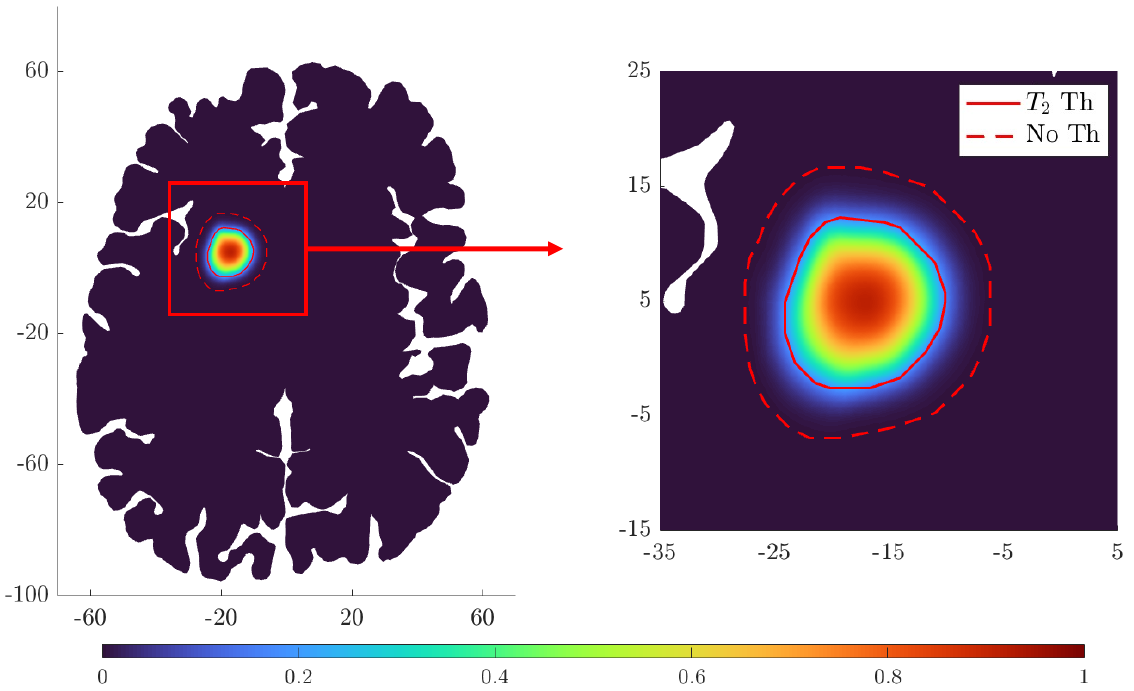}
       \caption{{\bf Tumor area contour.} Identification of the tumor-occupied region in a brain slice with (solid red line) and without (dashed red line) application of the $T_2$ detection threshold. The tumor distribution results from a 12-week simulation of the model without treatment, starting from the initial conditions given in \eqref{IC_M}–\eqref{IC_Q}. The right plot shows a magnification of the main area affected by the tumor.}
       \label{fig:Test1_Contour}
\end{figure}

\begin{figure}[h!]
       \centering
       \includegraphics[width=\linewidth]{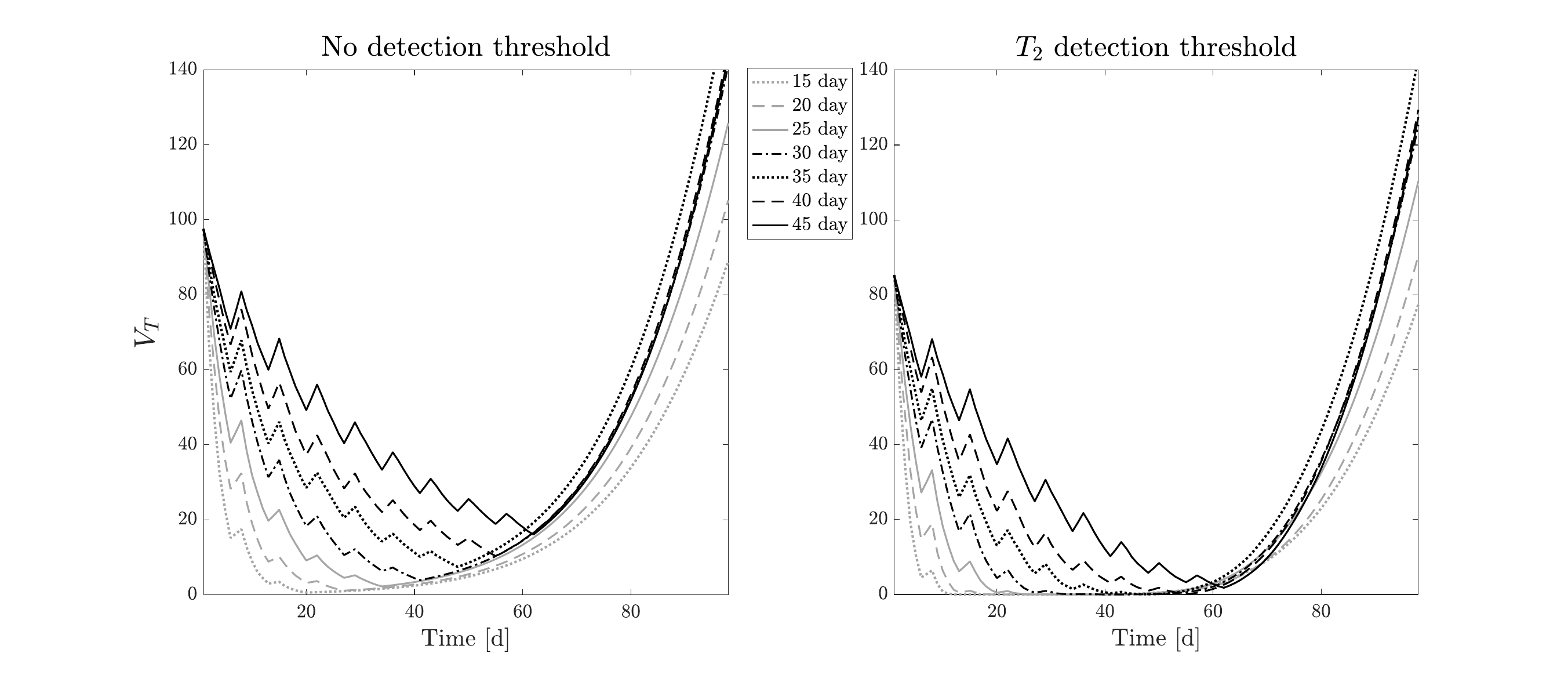}
       \caption{{\bf Impact of the detection threshold on tumor volume quantification.} Comparison of simulated mean tumor volume ($V_T$) over time with (right panel) and without (left panel) application of the $T_2$ detection threshold for the different treatment protocols. In the figure legend, the number $D_T$ of days for each treatment is indicated. Mean tumor volumes were obtained by averaging the results of 200 independent simulations of the same experiment.}
       \label{fig:Test1_Volume}
\end{figure}

To this aim, we use the tumor volume ($V_T$) as the primary response metric, comparing its evolution across the different treatment protocols over a total period of 14 weeks. This time frame includes both the treatment phase (ranging from 3 weeks for $D_T=15$ to 9 weeks for $D_T=45$) and the subsequent post-treatment rest period (correspondingly ranging from 11 to 5 weeks). {At each time point, the tumor volume is evaluated within the PTV by integrating the simulated tumor cell density. We consider two ways of processing the same simulated tumor distribution: (i) the total tumor burden, obtained by integrating the entire tumor cell density over the PTV; and (ii) the MRI-detectable tumor burden, obtained by restricting the integration to regions where the density exceeds 16\% of the carrying capacity, corresponding to the adopted $T_2$-Gd MRI detection threshold. Thus, the two measures differ in the detection criterion applied to the same underlying tumor distribution.} Figure~\ref{fig:Test1_Contour} illustrates an example of these two cases, showing the tumor region identified with a $T_2$ threshold (solid red line) and without a threshold (dashed red line), for a simulation in which the tumor evolved without treatment for 12 weeks.
Results of the tumor volume evolution over time, without (left panel) and with (right panel) application of the $T_2$-Gd MRI detection threshold for the different treatment protocols are shown in Figure \ref{fig:Test1_Volume}. 
From this figure, it is evident that for most treatment protocols ($D_T = 15, 20, 25, 30$), applying the $T_2$ detection threshold not only underestimates the actual tumor volume, but also suggests a transient complete remission ($V_T \approx 0$) lasting several weeks after therapy. In contrast, the results without threshold application reveal that a small residual tumor population survives in all cases, indicating that complete eradication is not achieved under any of the tested protocols. Specifically, in the $T_2$-threshold simulations, treatments with $D_T = 15, 20, 25,$ and $30$ show a temporary remission period of 1–4 weeks, followed by renewed tumor growth. When the full tumor mass is considered (no threshold), all treatment schedules exhibit eventual regrowth after therapy cessation. For completeness, in the Supplementary Figures~A.1 and A.2, we report the evolution of the mean tumor volume together with the minimum and maximum values obtained from 200 independent simulations of the same experiment for all analyzed treatment protocols. Analyzing the overall effect of the different treatment schedules, the longest delay in tumor recurrence is observed for $D_T = 15$ and $D_T = 20$, whereas protocols with $D_T = 30, 40,$ and $45$ produce comparable, yet less favorable outcomes. 

\subsection{Assessment of therapy efficacy: TCP/NTCP and R-score analysis}\label{subsec:Rscore}

\begin{figure}[h!]
       \centering
       \includegraphics[width=\linewidth]{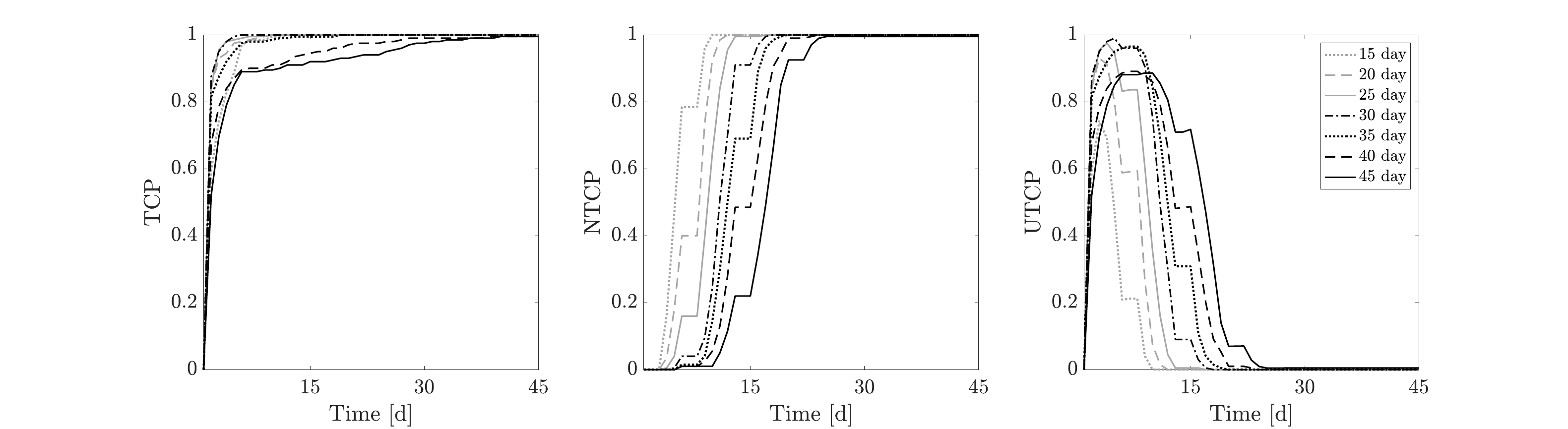}
       \caption{{\bf TCP, NTCP, and UTCP.} Comparison of the mean TCP (left panel), NTCP (middle panel), and UTCP (right panel) evolution over time for the different treatment protocols. In the figure legend, the number of $D_T$ for each treatment is indicated. Mean values for TCP, NTCP, and UTCP were obtained by averaging the results of 200 independent simulations of the same experiment.}
       \label{fig:Test1_TCP}
\end{figure}

To enable a more robust and reliable interpretation of the tumor volume analysis presented in Section~\ref{subsec:TV}, the results must be considered alongside the potential toxicity associated with each fractionation scheme~\cite{abou2004theoretical}. Accordingly, to evaluate the therapeutic trade-offs between tumor control and normal tissue damage, we employ three standard radiobiological metrics: tumor control probability (TCP), normal tissue complication probability (NTCP), and uncomplicated tumor control probability (UTCP) (see, e.g.,~\cite{McDermott2016}). Radiation therapy, in fact, represents a double-edged sword: while it effectively damages tumor cells, it inevitably affects normal tissue as well - albeit to a lesser extent. The ultimate goal is therefore to identify treatment protocols that maximize TCP while minimizing NTCP, thereby achieving high therapeutic efficacy with limited toxicity. These metrics are defined as follows.
\begin{itemize}
\item[$\bullet$] The tumor control probability quantifies the likelihood of complete tumor eradication and can be defined as 
\[
\text{TCP}(t):=\dfrac{\#~~\text{\rm (simulations with } \tau_M\le t)}{~~\text{\rm total }\# ~~\text{\rm simulations}}
\] 
where $\tau_M$ denotes the first time point at which the cancer cell density $M$ falls below 2\% of the carrying capacity across the entire PTV region.
\item[$\bullet$] The normal tissue complication probability estimates the probability of radiation-induced normal tissue injury and can be defined as 
\[
\text{NTCP}(t):=\dfrac{ \# ~~\text{\rm simulations with } \tau_Q\le t}{~~\text{\rm total } \# ~~\text{\rm simulations}}
\]
where $\tau_Q$ is defined as the first time point at which the healthy tissue density $Q$ decreases below 30\% of its original value within a subregion of the PTV of non-zero measure.
\item[$\bullet$] The uncomplicated tumor control probability combines the previous two metrics to describe the probability of achieving effective tumor control without severe normal tissue complications, and is defined as
\[
\text{UTCP}(t):=\text{TCP}(t)[1-\text{NTCP}(t)]
\] 
assuming independence between TCP and NTCP.
\end{itemize}
The results of these metrics for the different treatment protocols are presented in Figure \ref{fig:Test1_TCP}.
From this figure, we observe that the tumor response across the different treatment protocols, in terms of TCP, is quite similar and shows a high probability of tumor eradication within a relatively short time window (approximately 0–15 days), except for the hyper-fractionated schedules corresponding to $D_T = 40$ and $D_T = 45$. However, the corresponding NTCP results indicate that the hypo-fractionated treatments with $D_T = 15$ and $D_T = 20$ lead to {earlier onset of normal tissue damage and higher early-phase complication probability.} These findings should be interpreted in light of the results presented in Figure \ref{fig:Test1_Volume}, where $D_T = 15$ and $D_T = 20$ achieved the greatest reduction in tumor volume, albeit at the cost of {earlier and faster} normal tissue toxicity, as reflected by the NTCP evolution. When considering both tumor control and toxicity together through the UTCP metric (Figure \ref{fig:Test1_TCP}, right panel), intermediate fractionation schedules, such as $D_T = 25$, $D_T =30$, or $D_T = 35$, emerge as the most balanced options, providing substantial tumor control while limiting adverse effects on normal tissue. For completeness, the temporal evolution of the mean TCP values, along with their corresponding standard deviations for all treatment protocols, is presented in the Supplementary Figure~A.3.

\begin{figure}[h!]
       \centering
       \includegraphics[width=.5\linewidth]{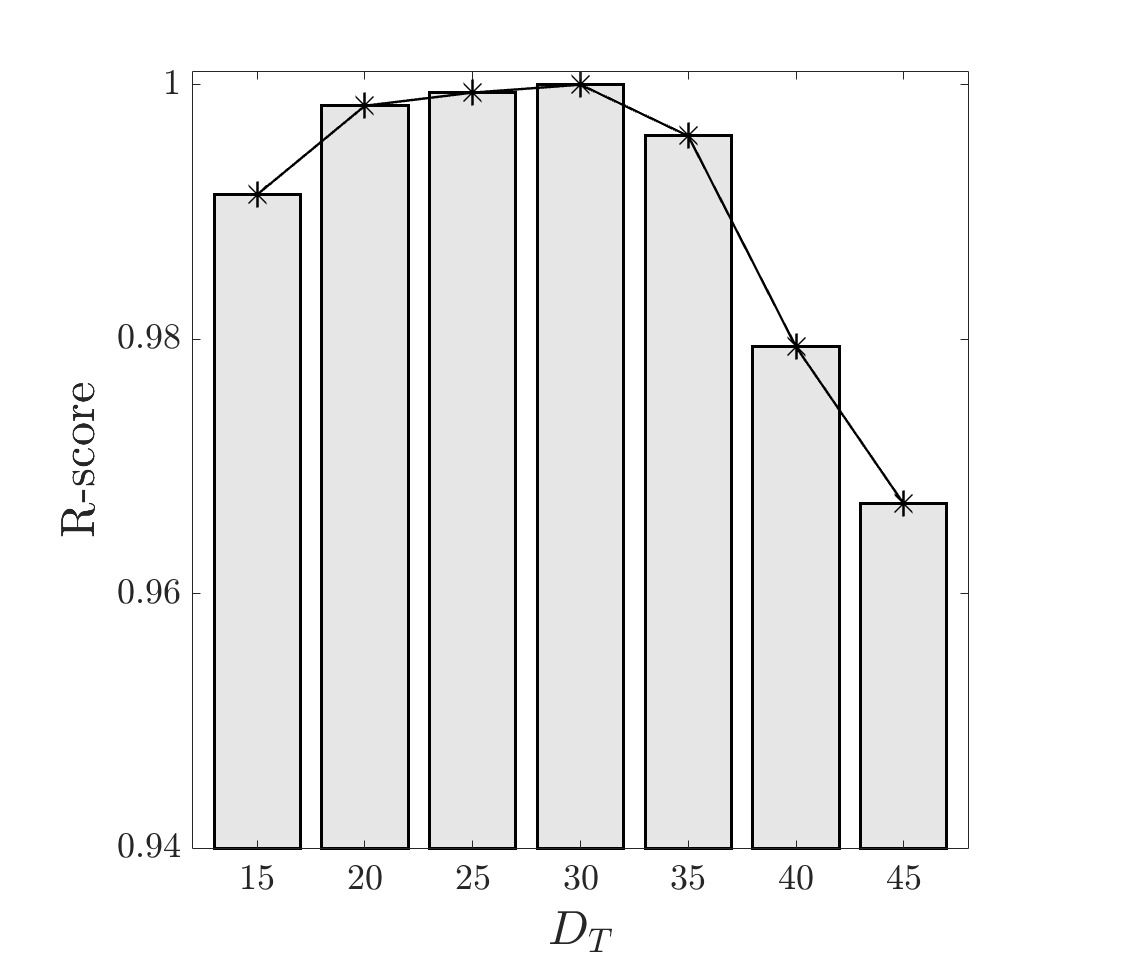}
       \caption{{\bf $R$-score analysis.} {Normalized $R$-score values for the different treatment protocols, with $D_T$ values shown on the x-axis.}}
       \label{fig:Rscore}
\end{figure}

The UTCP metric has been adopted by several authors to rank treatment plans \cite{leibel1991improved,chaikh2016should}. However, as pointed out in \cite{langer1998test}, where it is referred to as the $S$-score, this metric exhibits some undesirable properties, {such as an excessive sensitivity to (large) absolute NTCP values. Consequently, the ranking produced by UTCP is not invariant to a common rescaling of NTCP values, even when the relative differences between the treatment plans remain unchanged. Thus, the choice of the absolute scale of NTCP can influence treatment-plan rankings, which represents an important limitation of the metric.} To address this limitation, 
Brenner \cite{brenner1999more} proposed the $\mathcal {R}$-score index, defined as
\begin{equation}
{\mathcal {R\text{-score}}} = \frac{TCP}{NTCP}\,.
\end{equation}
In the present work, we adopt this metric to rank the proposed treatment plans. Specifically, since TCP and NTCP are defined as functions of time in our framework, we introduce a cumulative $R$-score, defined as
\begin{equation}
R\text{-score} = \sum_{t=1}^{45}\frac{TCP(t)}{NTCP(t)+\epsilon}\,.
\end{equation}
where $t$ is expressed in days ([d]). We modify the denominator of Brenner’s formula by introducing a small parameter $\epsilon$ to ensure mathematical correctness in cases where NTCP becomes zero. Since $\epsilon$ is assumed to be very small (specifically, we set $\epsilon=10^{-3}$), it does not affect the therapy assessment. To enable a relative comparison among treatment schedules rather than an assessment of their absolute efficacy, the $R$-score is normalized by the maximum achieved value. The resulting normalized index lies in the interval $[0,1]$, with higher values corresponding to more favorable treatment plans. Results of the $R$-score analysis are shown in Figure \ref{fig:Rscore}.

From Figure~\ref{fig:Rscore}, we observe that the treatments achieving the best outcome in terms of the $R$-score correspond to $D_T$ values between 20 and 35, with the optimal value obtained at $D_T=30$, which coincides with the standard therapy. However, the differences with settings relying on $D_T=20$ and $D_T=25$ are relatively small. This finding partially aligns with the results of the TCP/NTCP analysis and the tumor volume assessment. In particular, it appears that hyper-fractionated schedules yield lower $R$-scores, consistent with the observed length of tumor recurrence (see Figure~\ref{fig:Test1_Volume}), where the least favorable outcomes occur for $D_T=40$ and $D_T=45$. {To assess the robustness of these observations with respect to the level of stochastic variability, we additionally considered larger values of the stochastic diffusion parameter $\sigma_R$, up to $\sigma_R=0.85$. The resulting TCP and NTCP dynamics remained qualitatively consistent with those reported above, although the increased uncertainty led to greater variability in TCP and, on average, to a less efficient tumor eradication. We also observed a similar trend in the $R$-score analysis, namely a general decrease in the $R$-score value as $\sigma_R$ increased, with this effect being more pronounced for hyper-fractionated schedules characterized by smaller doses per delivery.}

\subsection{RECIST-based analysis of tumor response}
As a second numerical experiment, we assess the tumor response to the different treatment protocols using the Response Evaluation Criteria in Solid Tumors (RECIST) framework \cite{padhani2001recist,therasse2000new}. According to this classification, we define: complete response (CR) as the disappearance of all target lesions; partial response (PR) as a reduction of at least 65\% in the pre-treatment tumor volume; progressive disease (PD) as an increase of at least 73\% in the pre-treatment tumor volume; and stable disease (SD) as any change that does not meet the above criteria. {In addition to RECIST, the RANO criteria represent a standard framework for response assessment in neuro-oncology \cite{wen2023rano}. Here, we decide to adopt RECIST because its quantitative response categories can be directly applied to the simulated tumor volume.} 

\begin{figure}[h!]
       \centering
       \includegraphics[width=\linewidth]{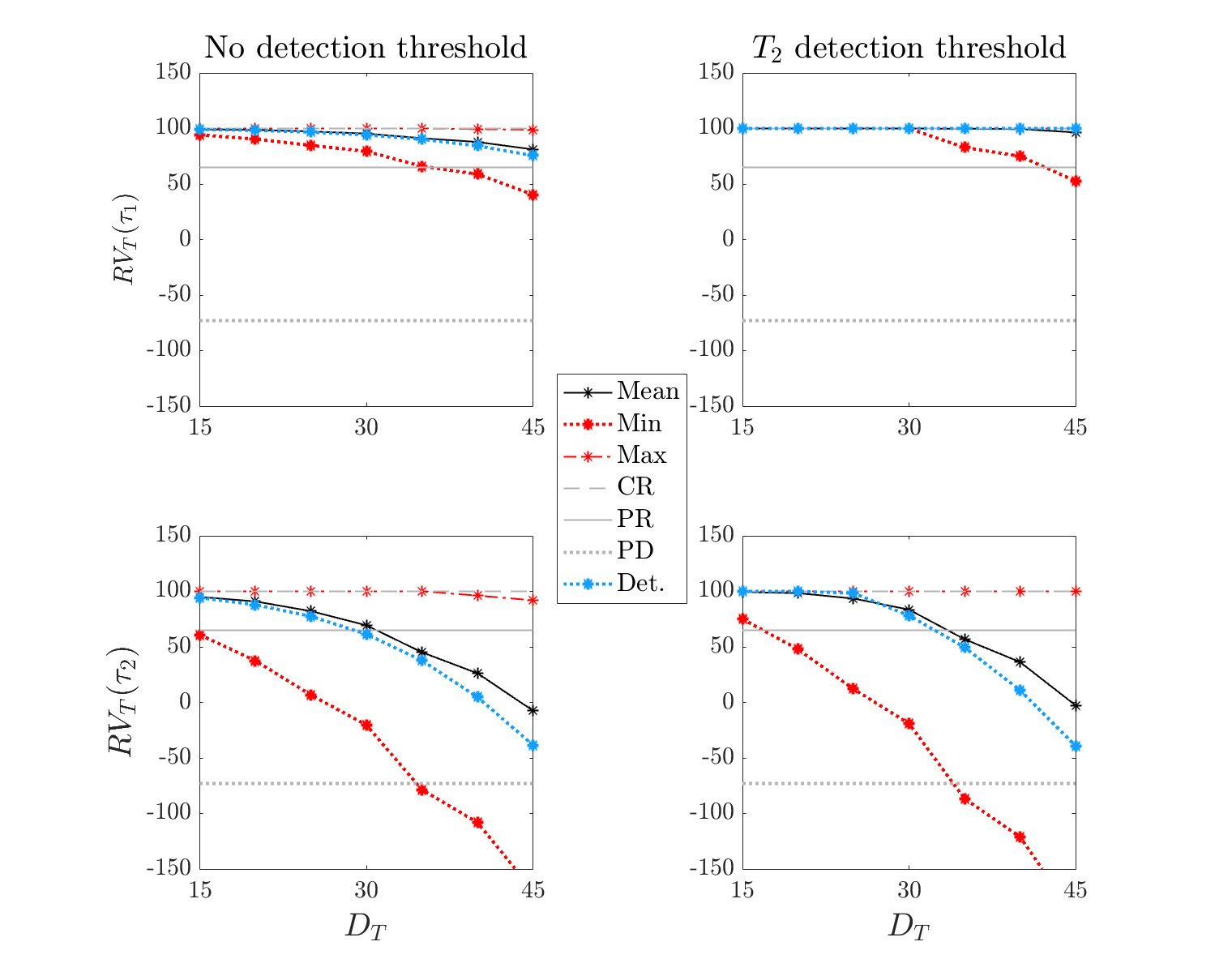}\\[-4ex]
       \caption{{\bf Treatment efficacy based on RECIST criteria.} Evolution of the percentage tumor volume reduction ($RV_T(\tau)$) as a function of $D_T$, evaluated at the end of the treatment ($\tau = \tau_1$, first row) and after the follow-up period ($\tau = \tau_2$, second row). Results are shown both without a detection threshold (left column) and with a $T_2$ detection threshold (right column). The mean percentage tumor volume reduction is shown (black line/dots), together with the minimum (red dotted line) and maximum (red dashed line) values obtained from 200 independent simulations. The dashed grey line indicates the complete response (CR) threshold; the region between the dashed and solid grey lines corresponds to partial response (PR); the region between the solid and dotted grey lines represents stable disease (SD); and values below the dotted line indicate progressive disease (PD). {The corresponding value of the tumor volume reduction obtained from the deterministic model ($\sigma_M=0$) for each of the treatment schedules is also shown for comparison (blue dotted line).}}
       \label{fig:Test2}
\end{figure}

Gliomas, due to their strong radioresistance and infiltrative growth, are typically classified clinically within the PD–PR range, with response primarily evaluated on T2-weighted images \cite{rockne2009mathematical}. Here, we quantify the tumor response using the percentage tumor volume reduction at time $t = \tau$, defined as
\[
RV_T(\tau):=\dfrac{(V_T(t=0)-V_T(t=\tau))}{V_T(t=0)}\times100,
\]
where $V_T$ denotes the total tumor volume.
We evaluate this metric at two distinct time points: $\tau = \tau_1$, corresponding to the last day of treatment, and $\tau = \tau_2$, corresponding to the follow-up period, defined as four weeks after the end of the treatment. For each case, $RV_T(\tau)$ is computed under two different assumptions: (i) without applying any detection threshold, and (ii) by applying a $T_2$ detection threshold for tumor volume estimation. The results are summarized in Figure \ref{fig:Test2}.
Specifically, Figure \ref{fig:Test2} compares the evolution of $RV_T(\tau)$ across the different treatment protocols, characterized by varying values of $D_T$, both without a detection threshold (left column) and with a $T_2$ detection threshold (right column). Results are shown at two time points: at the end of treatment ($\tau = \tau_1$, first row) and after the follow-up period ($\tau = \tau_2$, second row). Examining the results at the end of the treatment protocols, we observe that when the imaging detection threshold is applied, nearly all protocols (except for $D_T = 45$) appear, on average, to meet the CR criterion, indicating complete tumor eradication with a 100\% reduction in tumor volume after therapy. However, this apparent success is largely an artifact caused by the limited sensitivity of imaging in capturing the full tumor extent, as evident in the left panel of the first row of Figure \ref{fig:Test2}. In fact, when no detection threshold is applied, we note that from $D_T = 25$ onward, $RV_T$ values begin to decline, revealing that the tumor mass is not fully eradicated. Nonetheless, across all cases, the average response remains within the PR category, indicating a partial but consistent response to treatment. After the four-week follow-up period (second row of Figure \ref{fig:Test2}), the comparison between the left and right panels mirrors the previous findings: the use of a $T_2$ detection threshold results in a clear overestimation of treatment efficacy. Interestingly, despite this overestimation, all protocols show disease progression after follow-up, yet the mean response generally remains within the stable disease (SD) range. Only the minimum response values for $D_T = 35, 40,$ and $45$ fall into the PD region. Moreover, the standard therapeutic protocol ($D_T = 30$) consistently remains within the partial response (PR) range in both evaluation cases. These results are consistent with clinical observations of glioma evolution, which typically range between PD and PR classifications. {Finally, we compare the stochastic results with the corresponding deterministic model ($\sigma_M=0$). At the end of treatment ($\tau=\tau_1$), the mean tumor volume reductions predicted by the stochastic model are generally consistent with the deterministic results. Differences become more evident after the follow-up period, particularly for the hyper-fractionated treatment protocols. In these cases, the deterministic model generally predicts a greater degree of tumor progression than the mean stochastic response. This difference reflects the effect of the stochastic treatment response, whose magnitude is controlled by the diffusivity parameter $\sigma_M$ and can therefore modify the post-treatment tumor dynamics relative to the deterministic case.}

\subsection{Evaluation of the relapse-free probability }\label{KM_curves}
\begin{figure}[h!]
\centering
\includegraphics[width=.65\linewidth]{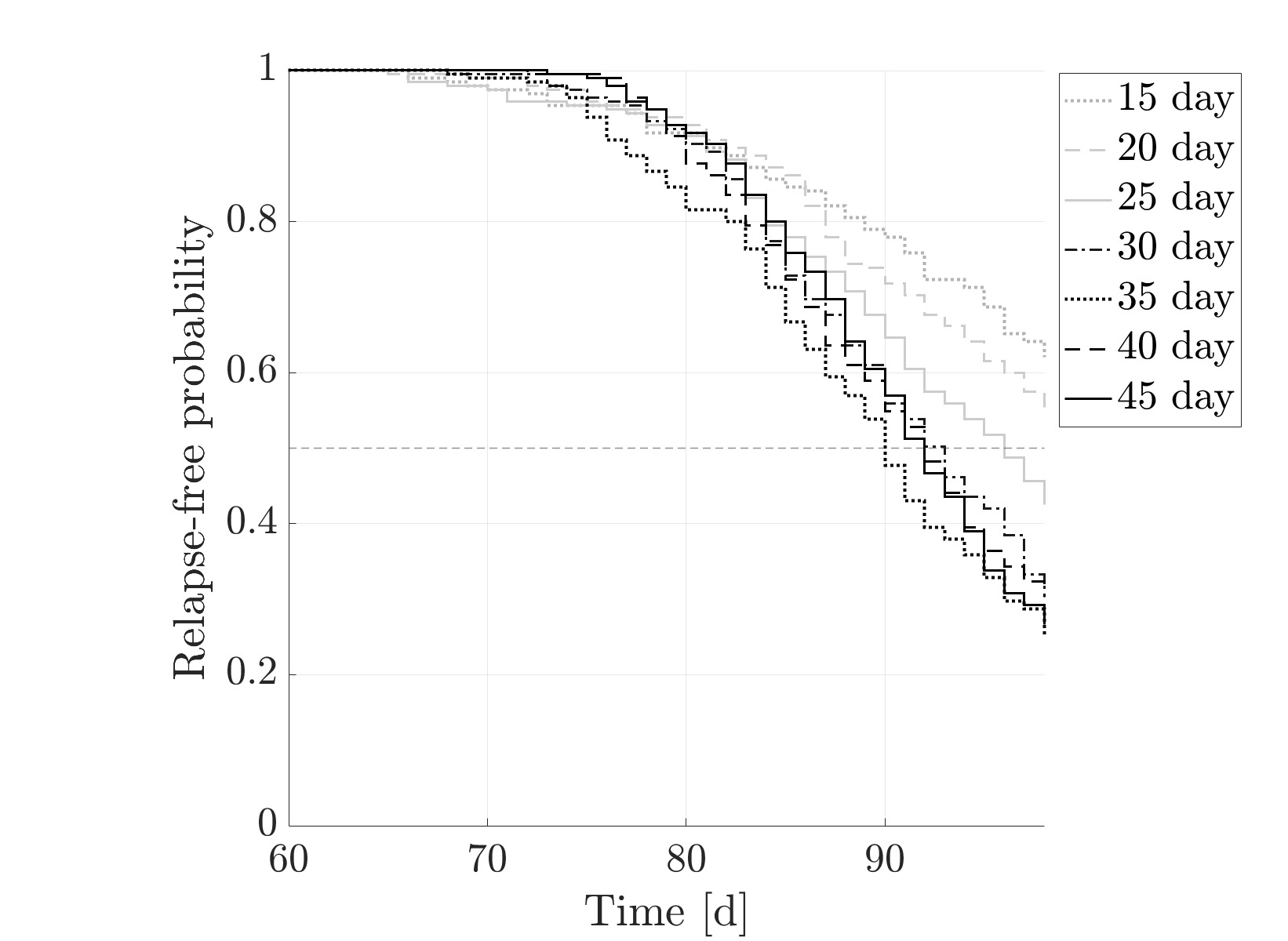}
\caption{{{\bf Kaplan-Meier curves.} Kaplan-Meier relapse-free survival curves for different treatment protocols ($D_T$ ranging from 15 to 45 days) over a 98-day window, displayed from day 60 onwards. The horizontal dashed grey line indicates the $50\%$ relapse-free probability. Curves are estimated from 200 independent stochastic realizations for each treatment protocol.}}
\label{fig:Test3}
\end{figure}

{As a final numerical experiment, we evaluate the tumor response to the different treatment protocols using the Kaplan-Meier estimator as a quantitative comparison metric. The Kaplan-Meier curves represent the probability of remaining relapse-free after the start of treatment as a function of time. For each treatment protocol, a relapse is considered the event of interest, occurring when the tumor volume becomes greater than or equal to its pre-treatment value. Simulations in which no relapse is observed within the 98-day observation period are treated as right-censored at the end of the observation window. The Kaplan-Meier estimator therefore accounts for both simulations in which relapse is observed and those in which the simulation ends before relapse occurs, providing a robust estimate of the probability of remaining relapse-free over the considered observation period~\cite{kaplan1958nonparametric}. The Kaplan-Meier curves for all treatment protocols are summarized in Figure~\ref{fig:Test3}, while the individual curves, together with their corresponding $95\%$ confidence intervals, are reported in the Supplementary Figure~A.4 for completeness.}

{As shown in Figure~\ref{fig:Test3}, a progressive variation in post-treatment relapse-free probability across the different protocols can be observed. In particular, the relapse-free probability decreases progressively as the treatment duration increases from $D_T=15$ to $D_T=30$ days, whereas these differences become less pronounced for longer treatment durations ($D_T>30$ days). Thus, within the considered 98-day observation window, shorter treatment protocols, {namely higher doses per fraction,} are associated with a higher probability of remaining relapse-free.}

{This result, however, should not be interpreted in isolation as evidence of better therapeutic efficacy of shorter treatment schedules. The relapse-free probability must be considered together with the treatment-related toxicity affecting healthy tissue, which is quantified through the UTCP analysis presented in Section~\ref{subsec:Rscore}. In particular, although shorter treatment schedules may provide a higher relapse-free probability within the considered observation window, they are also associated with an increased predicted risk of normal-tissue toxicity. Conversely, longer schedules may reduce treatment-related toxicity,  but provide a lower relapse-free probability. Among the protocols considered, the intermediate schedule with $D_T=25$ or $D_T=30$ days may provide a favorable balance between relapse-free control and normal-tissue toxicity, suggesting a potential compromise between treatment efficacy and tolerability within the framework of the present model.}

\section{Discussion}\label{sec:discussion} 
The model proposed in this work is designed to investigate the influence of different radiotherapy treatment protocols on glioma progression within the brain, with particular emphasis on how therapy-related randomness and imaging limitations can affect overall treatment outcomes. To our knowledge, this is the first continuous multiscale model that introduces stochasticity at the mesoscopic level in the therapy-related terms and then derives the corresponding random partial differential equations at the macroscopic scale, accounting for nonlinear, myopic self-diffusion and multiple taxis mechanisms. As such, it represents an advancement of the models presented in \cite{conte2020glioma,hunt2017multiscale}. However, the explicit incorporation of stochastic components constitutes the main novelty, enabling a randomness-aware—and therefore more realistic—interpretation of model predictions. Notably, our numerical experiments also employ evaluation metrics closer aligned with clinical practice than those typically used in mathematical oncology.

We started from the description of single-cell dynamics of glioma cells interacting with (macroscopic) healthy and cancer cell densities, with the aim of implementing both an anti-crowding mechanism and a haptotactic movement response towards increasing tissue densities. We formulated the corresponding kinetic transport equations for the glioma distribution function on the mesoscopic scale. At this stage, we included cell proliferation, triggered by cell receptor binding to tissue, as well as radiotherapy effects, which are modeled by means of a combination of the classical LQ radiobiological approach \cite{fowler1989linear} and an Ornstein–Uhlenbeck process to represent mean-reverting random fluctuations in treatment efficacy. By upscaling the mesoscopic transport equations in a parabolic manner, we formally deduced a population-level system for the evolution of the glioma cell density and the normal tissue density. The normal tissue was thereby assumed to be degraded by tumor cells and radiation treatment. The resulting macroscopic system is a set of coupled random partial differential equations featuring diffusion-advection-reaction terms in the tumor equation and ODE-type dynamics (no motility terms) for the space-time evolution of tissue.
Consistent with previous multiscale modeling works (e.g., \cite{conte_surulescu2020,conte2023mathematical,engwer2015glioma,engwer2016multiscale}), tumor migration is driven by terms carrying information from the microscopic scale. The myopic diffusion term involves a tumor diffusion tensor that explicitly accounts for the local tissue orientation. The haptotaxis term drives cells towards increasing tissue gradients, reflecting the cells' tendency to follow dense tissue structures. 
 
In the numerical experiments, we performed different tests aimed at addressing the tumor response to various treatment protocols using clinically-inspired settings and metrics. Specifically, we first utilized standard radiotherapy planning domains (i.e., GTV, CTV, and PTV) as reference regions. These domains reflect the radiological interpretation of imaging data and therewith associated detection thresholds, and they serve as the basis for evaluating tumor volume evolution over time. The detection threshold, which is related to imaging limitations, while not qualitatively changing the overall dynamics, leads to an underestimation of the tumor volume. More importantly, it introduces a temporal delay in identifying tumor relapse after the end of the treatment. Moreover, we assessed the tumor and normal tissue control probabilities (TCP and NTCP), enabling the exploration of therapeutic limitations related to the risk of severe normal tissue complications (NTCP). The use of these quantities as evaluation metrics demonstrated that intermediate fractionated treatment schedules (such as $D_T=25$, $D_T=30$, or $D_T=35$) achieve a better balance between tumor control and adverse effects. This result is partially confirmed by the $R$-score analysis, which enables us to rank the treatment schedules and indicates a better outcome for hypo-fractionated protocols compared to hyper-fractionated ones, with the best performance achieved by the standard schedule $D_T=30$. Furthermore, tumor response to therapy was also evaluated using the Response Evaluation Criteria in Solid Tumors (RECIST)  \cite{padhani2001recist}. The results obtained in this test were consistent with clinical observations of glioma evolution, which typically classify the response between progressive disease and partial response. {Finally, using Kaplan--Meier curves, we analyze the evolution of relapse-free probability across the different treatment protocols. The results indicate a more favorable response for shorter treatment protocols, while the relapse-free probability tends to show a similar evolution for longer treatment durations ($D_T > 30$), suggesting limited additional benefit from further prolonging the treatment.}

This work represents a first step toward a more detailed and realistic formulation of multiscale models that account for the effects of randomness in tumor outcomes. As such, it opens several directions for further development and future investigation. From a theoretical standpoint, some issues related to the existence and uniqueness of local solutions for the stochastic macroscopic system will be addressed in the forthcoming paper \cite{Hiremath2026}, although many questions remain open, for instance, those concerning the existence of global solutions for our {macroscale} system. 

{From a modeling perspective, an important future development will be to provide a more detailed representation of the different sources of uncertainty considered in the present framework. In particular, distinguishing between biological heterogeneity, such as spatial variations in cellular radiosensitivity and microenvironmental conditions, and technical or treatment-related uncertainties would allow their individual contributions to treatment response to be investigated more deeply. Similarly, incorporating spatial heterogeneity in the stochastic driver could provide a more realistic description of the local variability in radiation response. An additional important extension would be to account for realistic, spatially heterogeneous radiation dose distributions, referring to possible different dose-painting strategies.} A more detailed description of the evolution of tumor microenvironmental factors would also be essential. In particular, incorporating the dynamics of environmental acidity and hypoxia could provide further insight into the potential development of therapy resistance. {Moreover, accounting for patient-specific treatment characteristics, including the extent and timing of surgical resection, would allow the framework to be applied to a broader range of clinical scenarios and could contribute to a more comprehensive description of treatment response.} Finally, combining radiotherapy with other therapeutic approaches (such as chemo- or immuno-therapy), in which randomness also plays a significant role, would be of great interest. {Overall, these extensions could contribute to making the model progressively more comprehensive, ultimately supporting the investigation of patient-specific treatment responses and its potential translation toward clinical applications.}

\subsection*{Acknowledgment}
The authors gratefully acknowledge Hospital Galdakao-Usansolo (Galdakao, Spain) for providing the DTI dataset, as well as the clinicians Dr. Y. Dzierma (Rostock University Medical Center) and Dr. S. Knobe (Saarland University Medical Center) for their fruitful discussions and valuable suggestions. MC acknowledges support from the European Union - Next Generation EU, Mission 4, Component 1 (CUP: E13C24002380006) and the National Group of Mathematical Physics (GNFM-INdAM) (CUP E5324001950001). MC has also been partially supported by the State Research Agency of the Spanish Ministry of Science and FEDER-EU, project PID2022-137228OB-I00 (MICIU/AEI /10.13039/501100011033) and by Modeling Nature Research Unit, Grant QUAL21-011 funded by Consejería de Universidad, Investigaci\'on e Innovaci\'on (Junta de Andalucía). SAH acknowledges support from the Department of Mathematics at RPTU Kaiserslautern. This work was conducted in association with the CZS project \emph{AI-CARE}, which focuses on the development of PDE-based AI models and scRNA-seq data integration for discovering novel targets for GBM treatment.

\subsection*{Data availability statement}
{All relevant data are available within the manuscript and its Supporting Data. The code supporting the findings of this work, including the code required to reproduce the figures and plots, is publicly available at the following GitHub repository: \url{https://github.com/mconte93/GBM_StocastichModel_ConteEtAl_2026}} 

\subsection*{Competing interests}
The authors declare that the research was conducted in the absence of any commercial or financial relationships that could be construed as a potential conflict of interest.

\counterwithin{figure}{section}
\counterwithin{table}{section}
\counterwithin{equation}{section}
\renewcommand\thefigure{\thesection\arabic{figure}}
\renewcommand\thetable{\thesection\arabic{table}}
\renewcommand\thetable{\theequation\arabic{table}}

\phantomsection
\printbibliography
\end{document}